\pdfoutput=1
\documentclass[12pt,letterpaper]{article}
\usepackage{putex}
\usepackage{graphicx}
\usepackage{latexsym,amsmath,amsfonts,amssymb,amsthm}
\usepackage{bbm}
\usepackage{cite}
\usepackage{indentfirst}
\usepackage{booktabs,array}
\usepackage{placeins}
\usepackage{mathtools}
\usepackage{colonequals}
\usepackage{cancel}
\usepackage[mode=image|tex]{standalone}
\usepackage[dvipsnames]{xcolor}
\usepackage{tikz}
\usepackage{tikz-cd}
\usepackage{adjustbox}
\usepackage{genyoungtabtikz}
\usepackage{enumitem}
\usepackage{setspace}
\usetikzlibrary{decorations.markings}
\usepackage{graphicx}
\usepackage[margin=10pt,font=small,labelfont=bf]{caption}
\usepackage{subcaption}
\usepackage{xcolor}
\usepackage{float}
\theoremstyle{definition}
\newtheorem{claim}{Claim}
\usepackage{microtype}
\usepackage{hyperref}
\hypersetup{unicode}
\hypersetup{linktoc = all}
\hypersetup{pdfborderstyle={/S/U/W 0.5}}
\hypersetup{linkbordercolor = gray}
\hypersetup{bookmarksnumbered}
\usepackage[T1]{fontenc}
\usepackage[utf8]{inputenc}
\usepackage{lmodern}

\setlist{itemsep=2pt plus 1pt minus 1pt, topsep=2pt plus 1pt minus 1pt}

\renewenvironment{thebibliography}[1]{%
\begin{oldthebibliography}{#1}%
\setlength\itemsep{-2mm}%
}%
{%
\end{oldthebibliography}%
}

\newcommand{\eg}{\textsl{e.g.\@}}

\newcommand{\ie}{\textsl{i.e.\@}}

\numberwithin{equation}{section}

\newcommand{\nn}{\nonumber}

\DeclareMathOperator{\Tr}{Tr}

\DeclareMathOperator{\rank}{rank}

\DeclareMathOperator{\re}{\mathbb{R}e}
\DeclareMathOperator{\im}{\mathbb{I}m}

\DeclareMathOperator*{\SumInt}{%
\mathchoice%
  {\ooalign{$\displaystyle\sum$\cr\hidewidth$\displaystyle\int$\hidewidth\cr}}
  {\ooalign{\raisebox{.14\height}{\scalebox{.7}{$\textstyle\sum$}}\cr\hidewidth$\textstyle\int$\hidewidth\cr}}
  {\ooalign{\raisebox{.2\height}{\scalebox{.6}{$\scriptstyle\sum$}}\cr$\scriptstyle\int$\cr}}
  {\ooalign{\raisebox{.2\height}{\scalebox{.6}{$\scriptstyle\sum$}}\cr$\scriptstyle\int$\cr}}
}

\newcommand\qq{\mathbbmtt{Q}}

\begin{document}


\title{\begin{LARGE}
Schur correlation functions on $S^3\times S^1$\newline
\end{LARGE}}

\authors{Yiwen Pan$^1$ and Wolfger Peelaers$^2$
\medskip\medskip\medskip\medskip
 }

\institution{UU}{${}^1$
School of Physics, Sun Yat-Sen University, \cr
$\;\,$ Guangzhou, Guangdong, China}
\institution{Oxford}{${}^2$
Mathematical Institute, University of Oxford, Woodstock Road, \cr
$\;\,$ Oxford, OX2 6GG, United Kingdom}

\abstract{\begin{onehalfspace}{The Schur limit of the superconformal index of four-dimensional $\mathcal N=2$ superconformal field theories has been shown to equal the supercharacter of the vacuum module of their associated chiral algebra. Applying localization techniques to the theory suitably put on $S^3\times S^1$, we obtain a direct derivation of this fact. We also show that the localization computation can be extended to calculate correlation functions of a subset of local operators, namely of the so-called Schur operators. Such correlators correspond to insertions of chiral algebra fields in the trace-formula computing the supercharacter. As a by-product of our analysis, we show that the standard lore in the localization literature stating that only off-shell supersymmetrically closed observables are amenable to localization, is incomplete, and we demonstrate how insertions of fermionic operators can be incorporated in the computation.
}\end{onehalfspace}}

\preprint{}
\setcounter{page}{0}
\maketitle


{
\setcounter{tocdepth}{2}
\setlength\parskip{-0.7mm}
\tableofcontents
}


\section{Introduction}
In \cite{Beem:2013sza}, a remarkable, novel correspondence between four-dimensional superconformal field theories (SCFTs) with extended supersymmetry and chiral algebras was discovered.\footnote{\label{othercorrespondences}See \cite{Beem:2014kka,Chester:2014mea,Beem:2016cbd} for similar correspondences between six-dimensional maximally supersymmetric SCFTs and chiral algebras, and between three-dimensional $\mathcal N\geq 4$ superconformal field theories and one-dimensional topological algebras. The latter in fact define a deformation quantization of either the Coulomb or Higgs branch of vacua of the three-dimensional theory.} It arises by passing to the cohomology of either one of a pair of nilpotent supercharges, $\qq_1$ or $\qq_2$, in the superconformal algebra. Nontrivial cohomology classes at the origin of space are represented harmonically by a specific subset of supersymmetrically protected local operators. These operators precisely satisfy the shortening conditions that define states that contribute to the so-called Schur limit of the superconformal index \cite{Gadde:2011uv} and hence were named ``Schur operators'' in \cite{Beem:2013sza}. The cohomology classes can be transported in a two-dimensional plane while remaining in the kernel of either $\qq_i$. The generator responsible for the antiholomorphic coordinate dependence in this plane is $\qq_i$-exact, and as a result, within cohomology, one obtains holomorphic objects. The algebraic structure of these holomorphic cohomology classes descends from the operator product algebra of local operators in the full superconformal field theory and is that of a vertex operator algebra.\footnote{We use chiral algebra and vertex operator algebra interchangeably.}

Via this SCFT/chiral algebra correspondence, the vast landscape of four-dimensional SCFTs maps to a wealth of chiral algebras. A rich, ever-growing dictionary between quantities of the superconformal field theory and notions in the chiral algebra has been compiled. Moreover, the physics of four-dimensional SCFTs has served as inspiration for new understanding about and the construction of novel classes of vertex operator algebras \cite{Beem:2014rza,Lemos:2014lua,Nishinaka:2016hbw,Bonetti:2018fqz,Arakawa:2018egx},\footnote{Naturally, many four-dimensional SCFTs map to previously defined or constructed vertex operator algebras, see, \eg{}, \cite{Cordova:2015nma,Xie:2016evu,Song:2017oew,Buican:2017fiq,Choi:2017nur,Creutzig:2018lbc,Xie:2019yds}.} while, \textit{vice versa}, the rigid structure of chiral algebras is a powerful tool to gain new insights in SCFTs, see for example \cite{Buican:2015ina,Liendo:2015ofa,Cecotti:2015lab,Lemos:2015orc,Song:2016yfd,Buican:2016arp,Fredrickson:2017yka,Beem:2017ooy,Beem:2018duj}. 

In this paper we focus on a particular entry of the SCFT/chiral algebra dictionary. It is clear that the space of Schur operators defines the space of states of the vertex operator algebra, and thus it does not come as a surprise that the Schur limit of the superconformal index, $I(q;\vec a)$, which counts the former with signs, equals the (graded) vacuum character $\chi_0(q;\vec a)$ of the chiral algebra \cite{Beem:2013sza}:
\begin{equation}\label{indexequality}
\chi_0(q;\vec a)\colonequals\text{str}\ q^{L_0 - \frac{c_{2d}}{24}} \prod_i a_i^{J_0^{(i)}} =  q^{\frac{c_{4d}}{2}}\tr_{\mathcal H_{S^3}} (-1)^F q^{E-R} \prod_i a_i^{f_i}\equalscolon I(q;\vec a)\;.
\end{equation}
Here $\text{str}$ denotes the supertrace over the space of states of the chiral algebra, $L_0$ is the standard $\mathfrak{sl}(2)$ generator measuring holomorphic weight, and $c_{2d}$ is the central charge of the chiral algebra. On the right-hand side, the trace runs over the full Hilbert space $\mathcal H_{S^3}$ of states on the three-sphere of the four-dimensional SCFT, $F$ is the fermion number, $E$ denotes the conformal dimension while $R$ is the $SU(2)_R$ Cartan generator, and $c_{4d}$ is the trace anomaly coefficient associated with the square of the Weyl tensor.\footnote{Another entry in the SCFT/chiral algebra dictionary states that $c_{2d} = -12 c_{4d}$. The trace anomaly coefficient associated to the Euler density, $a_{4d}$, enters less directly in the chiral algebra. It controls certain representation theoretic properties of the chiral algebra \cite{Beem:2017ooy}.} Furthermore, recall from \cite{Beem:2013sza} that flavor symmetries in the four-dimensional theory manifest themselves as affine symmetries in the vertex operator algebra. For each Cartan generator $f_i$ of the four-dimensional flavor symmetry, we thus have an affine current $J^{(i)}(z)$ with modes $J_n^{(i)}$. Accordingly, we have decorated the trace formula of the Schur limit of the superconformal index with $a_i^{f_i}$, and the supertrace over the space of states of the chiral algebra with $a_i^{J_0^{(i)}}$. Thanks to cancellations between fermionic and bosonic states, only Schur states in $\mathcal H_{S^3}$ have a nonzero contribution to the trace in the right-hand side.

In this paper, our first aim is to shed light on the path integral origin of the above equality. Its right-hand side admits a path integral formulation as a partition function on a curved manifold that is topologically $S^3\times S^1$. We describe this background in detail and show that it preserves four supersymmetries that are organized in the superalgebra $\mathfrak{su}(1|1)\oplus \mathfrak{su}(1|1)$. Among them, one finds the counterparts of $\qq_1$ and $\qq_2$. Localization techniques \cite{Witten:1988ze,Witten:1991zz,Pestun:2007rz}  allow one to ``localize'' the path integral over the infinite-dimensional space of quantum fields to an integral over a simpler slice of field configurations. We will show that a judicious choice of localization supercharge $\mathcal Q$ precisely results in this slice describing the chiral algebra living on a torus inside $S^3\times S^1$.\footnote{Other localization computations in which the theory localizes onto a quantum field theory on a lower-dimensional manifold have been considered in \cite{Pestun:2009nn,Dedushenko:2016jxl,Dedushenko:2017avn,Bonetti:2016nma,Pan:2017zie,Mezei:2018url,Dedushenko:2018icp}.} To wit, we take $\mathcal Q = \qq_1 + \qq_2$. For a gauge theory with gauge group $G$, and associated Lie algebra $\mathfrak g$, whose conformal matter content is given by hypermultiplets transforming in some representation $\mathfrak R$ of the combined gauge and flavor group, the result of the localization computation takes the form
\begin{equation}\label{localizationresult}
I(q;\vec a) = \int[D\Phi] e^{-S[\Phi]} = \frac{1}{|W|} \oint \prod_{j=1}^{\rank \mathfrak g} \frac{db_j}{2\pi i b_j}\ Z_{\text{1-loop}}^{\text{VM}}(q,\vec b) \int[DQ D\widetilde Q] \ e^{- S^{T^2}_\text{SB}[Q, \widetilde Q]}\;.
\end{equation}
Here $\Phi$ collectively denotes the fields of the superconformal field theory on $S^3\times S^1$. Furthermore, $|W|$ denotes the order of the Weyl group of $\mathfrak g$. The contour integration on the right-hand side is over the maximal torus of the gauge group. Its integrand contains the one-loop determinant of quadratic fluctuations of the vector multiplet and a path integral over a dynamically and background gauged symplectic boson pair $Q,\widetilde Q$. The fields $Q,\widetilde Q$ transform in representations $\mathfrak R$ and $\mathfrak R^*$ respectively, and live on the above-mentioned torus inside $S^3\times S^1$. Their dynamics is described by the quadratic action $S^{T^2}_\text{SB}[Q, \widetilde Q] = \frac{1}{\pi \ell} \int  \sqrt{g}d^2z  \, \widetilde Q\, D_{\bar z} Q $. The complex structure parameter $\tau$ of said torus is related to $q$ in the standard fashion: $q=e^{2\pi i \tau}$. The path integral over $Q, \widetilde Q$ can be performed explicitly and reproduces the standard hypermultiplet contribution $Z^\text{HM}(q,\vec b,\vec a)$ to the matrix integral computing the superconformal index.

The supertrace over the space of states in the left-hand side of \eqref{indexequality} can be further decorated with the insertion of chiral algebra fields. If the inner product between states is diagonal in conformal weight, the supertrace will pick out the overall zero-mode of the insertion. Similarly, one can insert additional (transported) Schur operators in the superconformal index -- after all, local operators act as endomorphisms on the Hilbert space. Our localization computation can be generalized to include such insertions. This is a nontrivial and non-obvious statement for two related reasons. The standard lore states that a localization computation can be enriched with insertions of off-shell $\mathcal Q$-closed bosonic observables.\footnote{We are considering localization computations that do not suffer from fermionic zero-modes. If they are present, the integrations over these modes of course need to be saturated by the appropriate number of fermionic insertions.} Indeed, the off-shell closedness of the observable is typically put forward as a condition that, when satisfied, allows one to deform the action by a simultaneously $\mathcal Q$-exact and $\mathcal Q$-closed term without changing the result of the path integral. Choosing an appropriate such deformation term and sending the deformation parameter $\mathfrak s$ to infinity, justifies to perform a WKB approximation that happens to compute exact results. The restriction to bosonic observables stems from the observation that, naively, fermionic insertions would be evaluated on their vanishing classical values. The correlators we are after are, however, neither closed off-shell nor purely bosonic. In section \ref{section:insertions}, we will revisit the localization logic and will show how these naive issues can be overcome. In short, the condition for independence of the deformation parameter $\mathfrak s$ in the presence of an insertion is a full-blown path integral statement. Moreover, the observable itself can depend on the deformation parameter, and in practice we indeed find that each fermionic letter in the observable $\mathcal O_{(\mathfrak s)}$ should be accompanied by a factor $\sqrt{\mathfrak{s}}$. In our localization computation on $S^3\times S^1$, we show by explicitly performing the path subintegral over the auxiliary fields in the condition for independence, that if the operator $\mathcal O_{(\mathfrak s)}$ is closed on-shell, \ie{}, if it is annihilated by the supersymmetry variations obtained from those of $\mathcal Q$ by substituting the auxiliary field equations, then it can be localized. We remind the reader that the condition of on-shell closedness of gauge invariant operators precisely coincides with the condition that ensures that a Schur operator remains Schur after performing an exactly marginal gauging, see \cite{Beem:2013sza}, implying that all ($S^3\times S^1$ versions of) transported Schur operators are in fact ``localizable.'' The fermionic nature of the insertion is no obstruction, since the localization is only active in the bosonic sector. The result of the localization computation in the presence of (a product of) insertions, $\mathcal O$, which are necessarily inserted on the torus $T^2\subset S^3\times S^1$, is
\begin{equation}
\langle \mathcal O \rangle  = \frac{1}{I(q;\vec a)}\frac{1}{|W|} \oint   \prod_{j=1}^{\rank \mathfrak{g}} \frac{db_j}{2\pi i b_j}\; Z_\text{1-loop}^\text{VM}(q,\vec b)\; Z^\text{HM}(q,\vec b,\vec a) \; \langle \mathcal{O} \rangle_{\text{GT}} \ ,
\end{equation}
where $\langle \ldots \rangle_{\text{GT}}$ denotes the evaluation of the correlator in the Gaussian theories described by the propagators reported in the main text, see \eqref{fermionicprop} and \eqref{bosonicprop}. In other words, a simple evaluation using Wick's theorem suffices to compute $\langle \ldots \rangle_{\text{GT}}$. It is worth pointing out that the Gaussian theory describing the dynamics of $Q$ and $\widetilde Q$ is of course described by the above-encountered action $S^{T^2}_{\text{SB}}$, while the propagator for the fermionic letters $\lambda_z$ and $\tilde\lambda_z$ can be interpreted as defining that subsystem as the small $(b,c)$ ghost system, in line with the chiral algebra/SCFT dictionary.\footnote{The vector multiplet one-loop determinant can similarly be seen to correspond to the path integral over said small $(b,c)$ ghost system.}

The plan of the paper is as follows. In section \ref{section:SchurIndex+SuGra}, we recall the Schur limit of the superconformal index and present the corresponding supergravity background on $S^3\times S^1$. In section \ref{section:localizingSchur}, we perform the detailed localization computation of the Schur index. In section \ref{section:insertions} we study correlation functions of local insertions. In particular, we show how localization computations can be used to calculate correlation functions of non-off-shell closed operators and of fermionic observables. We conclude with a discussion and some future directions in section \ref{section:discussion}. Two appendices contain definitions and technical details of some computations. In particular, in appendix \ref{specialfuctions} we summarize the definitions and some useful properties of the Jacobi theta functions $\theta_i(z|\tau)$ and of the twisted Weierstrass function. In appendix \ref{app:Q-exactness} we study the $\mathcal N=2$ (anti)chiral supermultiplet, which we use to show that the (anti)holomorphic Yang-Mills lagrangian is $\mathcal{Q}$-exact and plays an important role in deriving localizability conditions.


\section{Schur index and its supergravity background}\label{section:SchurIndex+SuGra}
In this section, we recall the definition of the Schur limit of the superconformal index of four-dimensional $\mathcal N=2$ superconformal field theories (SCFTs) \cite{Gadde:2011uv}, often simply referred to as the Schur index. We also identify the (conformal) supergravity background with the property that the partition function of a four-dimensional SCFT placed thereon computes the Schur index. We construct the solutions to the generalized Killing spinor equations on this background. Finally, we recall the supersymmetric transformation rules and actions, and show that the Yang-Mills action is exact with respect to all supercharges preserved by the background. In particular, we thus prove that the partition function is independent of exactly marginal couplings.

\subsection{The Schur limit of the superconformal index}\label{subsection:schurlimitdefn}
The Schur index of a four-dimensional $\mathcal N=2$ superconformal field theory is defined as
\begin{equation}\label{SchurIndexDefn}
I(q,\vec a)\colonequals \Tr_{\mathcal{H}_{S^3}}\, (-1)^F\, q^{E-R}\, \prod_{j=1}^{\rank G_F} a_j^{f_j}\;,
\end{equation}
in terms of a trace over the Hilbert space $\mathcal{H}_{S^3}$ of states on the three-sphere. The trace contains $(-1)^F$, with $F$ the fermion number, a fugacity $q$ coupling to the difference of the conformal dimension $E$ and the $SU(2)_R$ $R$-symmetry Cartan generator $R$, and flavor fugacities $a_i$ associated with the Cartan generators of the flavor group $G_F$. As in the standard Witten index \cite{Witten:1982df}, the presence of $(-1)^F$ in the trace ensures vast cancellations between bosonic and fermionic states and guarantees that only certain states residing in short multiplets that cannot recombine into long ones, contribute\cite{Kinney:2005ej}. In the case at hand, only states for which both
\begin{alignat}{2}
&\Delta_1 \colonequals \{ Q_{1-}, S^{1-}\} &&= \frac{1}{2}\left(E -2R -\mathcal M\right) - \frac{1}{4}\left(r+ 2\mathcal M^\perp\right)\\
&\Delta_2 \colonequals \{ \widetilde S_2^{\dot -}, \widetilde Q^2_{\dot-} \}&& = \frac{1}{2}\left(E -2R -\mathcal M\right) + \frac{1}{4}\left(r+ 2\mathcal M^\perp\right)
\end{alignat}
vanish identically, contribute.\footnote{We denote the Poincar\'e supercharges as $Q_{I\alpha}, \widetilde Q^I_{\dot\alpha}$ and the special conformal charges as $S^{I\alpha},\widetilde S_I^{\dot\alpha}$. Here $I$ is an $SU(2)_R$ index and $\alpha,\dot\alpha$ are indices for the $SU(2)_1\times SU(2)_2 \simeq SO(4)$ rotational group. Hermitian conjugation (in radial quantization) acts as $(Q_{I\alpha})^\dagger = S^{I\alpha}, (\widetilde Q^I_{\dot\alpha})^\dagger = \widetilde S_I^{\dot\alpha}$. } Here $r$ is the $U(1)_r$ $r$-symmetry charge, and $\mathcal M$ and $\mathcal M^\perp$ are generators of the $SO(2)$ rotational groups of two orthogonal planes in $\mathbb R^4$.\footnote{In our conventions, $\mathcal M$ rotates the $(x_3,x_4)$-plane while $\mathcal M^\perp$ acts on the $(x_1,x_2)$-plane. In terms of the generators $\mathcal M_\alpha^{\phantom{\alpha}\beta}, \mathcal M^{\dot{\alpha}}_{\phantom{\alpha}\dot\beta} $ of the rotational group $SU(2)_1\times SU(2)_2 \simeq SO(4)$, they are identified as $\mathcal M \colonequals \mathcal M_+^{\phantom{+}+} + \mathcal M^{\dot +}_{\phantom{+}\dot+}$ and $\mathcal M^\perp \colonequals \mathcal M_+^{\phantom{+}+} - \mathcal M^{\dot +}_{\phantom{+}\dot+}$.} Equivalently, one may characterize states contributing to the index as those annihilated by the two Poincar\'e supercharges $Q_{1-},\widetilde Q^2_{\dot-} $ and their Hermitian conjugates. Note that $\Delta_1, Q_{1-}, S^{1-}$ and $\Delta_2 ,\widetilde S_2^{\dot -}, \widetilde Q^2_{\dot-}$ define two commuting $\mathfrak{su}(1|1)$ algebras. It is also important to keep in mind that the trace implicitly contains $e^{-\beta_1 \Delta_1}\, e^{-\beta_2 \Delta_2}$, but is ultimately independent of both $\beta_1$ and $\beta_2$. 

On all states in the theory, as a consequence of unitarity, $\Delta_1 \geq 0$ and $\Delta_2 \geq 0$. Hence, the condition that states be annihilated by both $\Delta_1$ and $\Delta_2$ is equivalent to the single requirement that $E -2R -\mathcal M=0$. States satisfying this latter condition are called Schur states. They define harmonic representatives of cohomology classes of the nilpotent supercharge $\qq_1 =  Q_{1-} + \widetilde S_2^{\dot -}$ or, equivalently, the charge $\qq_2 = S^{1-} - \widetilde Q^2_{\dot-}$.\footnote{One can in fact define various other, inequivalent nilpotent supercharges whose harmonic representatives are characterized by the condition $E -2R -\mathcal M=0$. Those supercharges will not play a role here.} Indeed,
\begin{equation}
\qq_1^2 = \qq_2^2 = 0\;, \qquad \{\qq_1,\qq_1^\dagger\} = \{\qq_2,\qq_2^\dagger\} = E -2R -\mathcal M\;.
\end{equation}
In \cite{Beem:2013sza}, it was shown that cohomology classes of $\qq_i$, for either $i=1,2$, of local operators inserted away from the origin are endowed with the structure of a chiral algebra.\footnote{Recall that states of a CFT are in one-to-one correspondence to operators inserted at the origin through the standard state-operator correspondence.} This algebraic structure descends from the operator product algebra of local operators of the four-dimensional SCFT. As an immediate consequence, the space of (cohomology classes of) states contributing to the Schur index is isomorphic to the space of states of said chiral algebra and the Schur index equals the (graded) vacuum character or torus partition function of the associated chiral algebra. The first goal of this paper is to elucidate this latter equality directly from the path integral.

\subsection{Supergravity background}\label{subsection:sugra_background}
A systematic approach to place theories on a curved background was pioneered in \cite{Festuccia:2011ws}. It amounts to defining a profile for the bosonic fields in the off-shell supergravity multiplet and solving the equations setting to zero the variations of its fermionic fields. The latter solutions are in one-to-one correspondence to rigid supersymmetries preserved on the curved background. To compute the Schur index as a partition function, our first task is thus to specify the relevant supergravity background. Concretely, we will consider an $\mathcal N=2$ conformal supergravity background, as for example also used in \cite{Hama:2012bg,Pestun:2014mja}.

\begin{figure}
  \centering
	\includegraphics[width=0.35\textwidth]{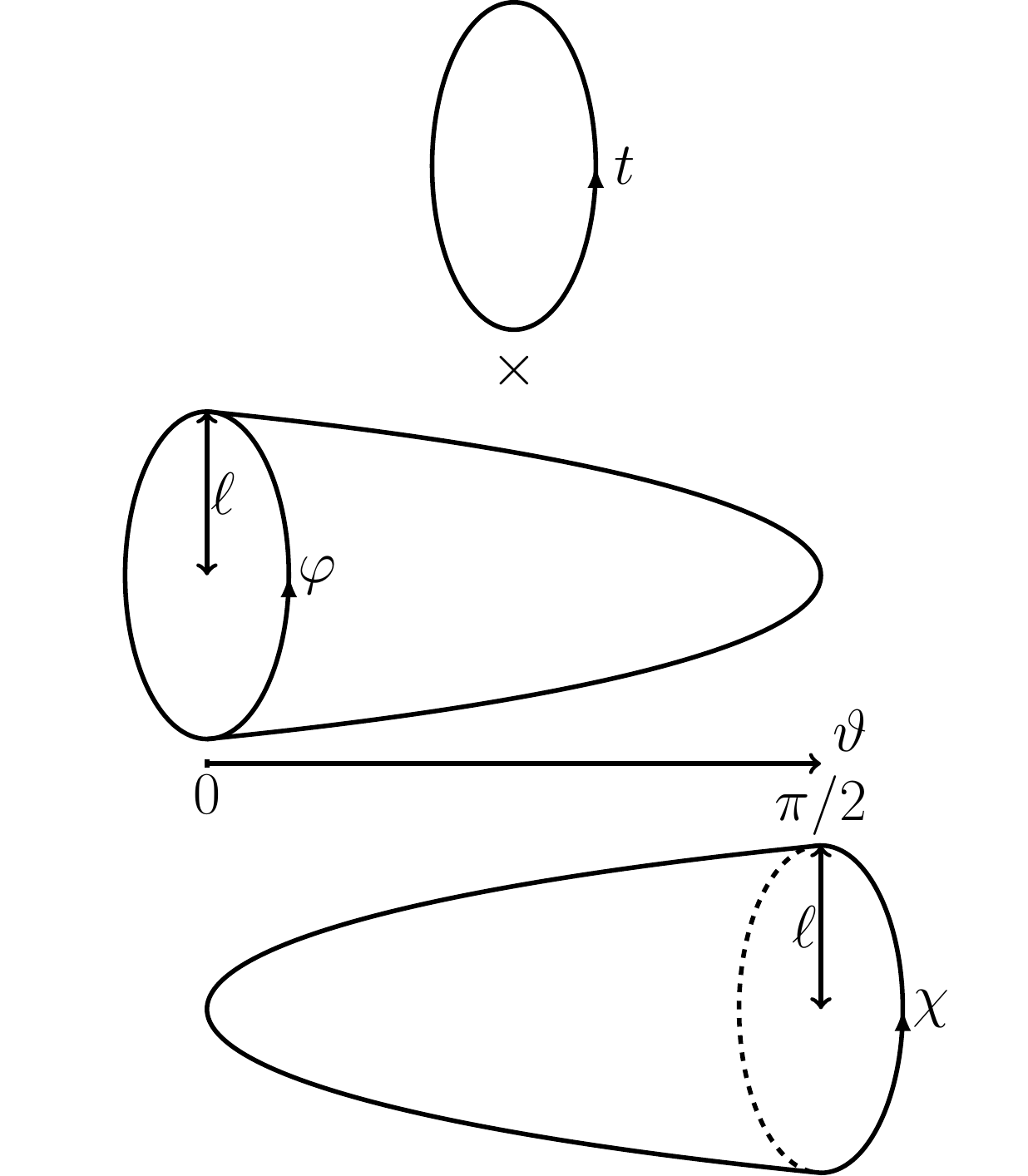}
	  \caption{Depiction of $S^3\times S^1$ and its coordinates. The three-sphere is represented as a torus-fibration over an interval.\label{fig:S3xS1}}
\end{figure}

The geometric background is described by the metric
\begin{multline}
ds^2 = \ell^2 \Big(\cos^2\vartheta \big(d\varphi-\frac{i}{4\pi}(\beta_1+\beta_2) dt\big)^2 + \sin^2\vartheta \big(d\chi-\frac{i}{4\pi}(\beta_1-\beta_2) dt\big)^2 + d\vartheta^2\Big)\\ + \big(-i \ell \tau +\frac{\ell}{4\pi}(\beta_1+\beta_2)\big)^2 dt^2\;.
\end{multline}
Here the coordinates $\varphi,\chi$ and $t$ are periodic with period $2\pi$ and $\vartheta \in [0,\frac{\pi}{2}]$. The metric describes an $S^3$-fibration over $S^1$, where in turn the three-sphere of radius $\ell$ is described as a torus-fibration over the $\vartheta$-interval. Finally, we conveniently parametrized the fugacity $q$ as 
\begin{equation}\label{}
q\colonequals e^{2\pi i \tau}\;.
\end{equation}
Figure \ref{fig:S3xS1} clarifies the definition of the coordinates. 
The background also turns on a flat $SU(2)_R$ and $U(1)_r$ connection
\begin{equation}\label{SU2RandU1rconnection}
(V_t^{(R)})_I^{\phantom{J}J} = -\frac{1}{2} \big(\tau + \frac{i}{2\pi}(\beta_1+\beta_2)\big)\begin{pmatrix}
1 & 0 \\
0 & -1
\end{pmatrix}_I^{\phantom{J}J}\;,\qquad\qquad V_t^{(r)} = -\frac{i}{8\pi}(\beta_1-\beta_2)\;.
\end{equation}
The final nonzero field in the supergravity background is a scalar field
\begin{equation}
  M = - \frac{2}{\ell^2}\;.
\end{equation}

The inclusion of the flavor fugacities $a_i$ in the trace corresponds to turning on a flat connection for background vector multiplets for the Cartan subgroup $U(1)^{\rank G_F}$ of the flavor symmetry group $G_F$
\begin{equation}\label{flavorbg}
A_j^{\text{bg}} = \mathfrak{a}_j dt \;,
\end{equation}
where we parametrized $a_j \colonequals \exp(2\pi i \mathfrak a_j)$.

The supergravity background depends on the parameters $\beta_1$ and $\beta_2$ coupling to the charges $\Delta_1$ and $\Delta_2$. From the point of view of the index, it is clear that both $\beta$s are immaterial and can be chosen arbitrarily.\footnote{From the point of view of the partition function on the supergravity background, the parameters $\beta_1$ and $\beta_2$ will correspond to supersymmetrically exact deformations.} We can choose to make use of this freedom to simplify the background. In particular, it is sometimes useful to ensure that the metric is real by choosing $\beta_1$ and $\beta_2$ purely imaginary and such that $\re{\tau} = -\frac{i}{4\pi}(\beta_1+\beta_2)$. We will take the $\beta$s imaginary throughout the paper, so that in particular the $U(1)_r$ connection is real, but will not impose the equality on the real part of $\tau$ unless stated explicitly.

We choose the vielbein
\begin{equation}\label{vielbein}
\begin{aligned}
&e^1=\ell\cos\vartheta\big(d\varphi-\frac{i}{4\pi}(\beta_1+\beta_2) dt\big)\;, \qquad && e^2=\ell \sin\vartheta \big(d\chi-\frac{i}{4\pi}(\beta_1-\beta_2) dt\big)\;,\\
&e^3=\ell d\vartheta\;, \qquad && e^4=\big(-i \ell \tau +\frac{\ell}{4\pi}(\beta_1+\beta_2)\big) dt\;.
\end{aligned}
\end{equation}
The nonzero components of the spin-connection are
\begin{equation}
\omega^{13}=-\omega^{31}=-\sin\vartheta\big(d\varphi-\frac{i}{4\pi}(\beta_1+\beta_2) dt\big)\;, \qquad \omega^{23}=-\omega^{32}=\cos\vartheta \big(d\chi-\frac{i}{4\pi}(\beta_1-\beta_2) dt\big)\;.
\end{equation}

The locus $\vartheta=0$ defines a torus with coordinates $\varphi, t$ and will play a central role in our computation; we denote it as $T^2_{\vartheta=0}$ or simply $T^2$. This torus can be parametrized by the coordinates
\begin{equation}
z\colonequals \varphi + \tau t\;, \qquad \tilde z\colonequals \varphi + \tilde \tau t\;,
\end{equation}
where $\tilde\tau$ is defined as
\begin{equation}
\tilde \tau \colonequals -\tau - \frac{i}{2\pi}(\beta_1+\beta_2)
\end{equation}
The coordinates $z$ and $\tilde z$ satisfy the obvious periodicity conditions
\begin{equation}\label{periodicitycoords}
z \sim z + 2\pi (m + n \tau)\;, \qquad \tilde z \sim \tilde z + 2\pi (m + n \tilde \tau)\;, \qquad \text{with}\qquad m,n\in \mathbb Z\;.
\end{equation}
In these coordinates, the metric on $S^3 \times S^1$ reduces to a metric on $T^2_{\vartheta=0}$, which takes the form
\begin{equation}\label{induced-torus-metric}
\widehat{ds^2}_{T^2_{\vartheta=0}} = \ell^2 dz d\tilde z\;,
\end{equation}
and correspondingly, we denote the metric components as $(\widehat{g_{T^2}})_{\hat\mu\hat\nu}$. Note that for generic values of $\beta_1$ and $\beta_2$, this metric is \emph{not} K\"ahler with respect to the usual complex structure associated to $\tau$; however, when the $\beta$s are chosen purely imaginary and such that $\re{\tau} = -\frac{i}{4\pi}(\beta_1+\beta_2)$, the metric $\widehat{ds^2}_{T^2_{\vartheta=0}}$ does become K\"ahler, with $\ell$ the K\"ahler parameter. Indeed, for these values of the $\beta$s, we easily find that $\tilde \tau = \bar \tau$ and $\tilde z = \bar z$. In the next section, we will argue that the final result of the localization computation does not depend on $\beta_1$, $\beta_2$, hence, without loss of generality, we can choose this K\"ahler limit from that point onwards. In fact, from that point onwards, we choose to supplement the K\"ahler limit with $\beta_1=\beta_2$.\footnote{This additional choice for the $\beta$-parameters is useful to eliminate the $U(1)_r$ connection.} We will denote the K\"ahler metric by $ds^2$. It is given concretely by $ds^2 = \ell^2 |d\varphi + \tau dt|^2 = \ell^2 dz d\bar z$. Its volume form reads $\vol(T^2_{\vartheta=0})= \sqrt{g}\ dz\wedge d\bar z = \ell^2 \im\tau\ d\varphi\wedge dt$, where $\sqrt{g} = \frac{i\ell^2}{2}$.

\subsection{Killing spinors}
The background defined in the previous subsection admits four solutions to the generalized Killing spinor equations \cite{Hama:2012bg,Pestun:2014mja}
\begin{equation}\label{KSE}
D_m \xi_I = -i \sigma_m \tilde \xi_I'\;, \qquad D_m \tilde\xi_I = -i \tilde\sigma_m \xi_I' \;,
\end{equation}
for some arbitrary primed spinors. Here we have omitted the supergravity fields that are not turned on in our background. The covariant derivatives include the $SU(2)_R$ and $U(1)_r$ connection:
\begin{equation}
D_m\xi_I = \partial_m \xi_I + \frac{1}{4}\omega_m^{ab}\sigma_{ab} \xi_I - i (V_m^{(R)})_I^{\phantom{J}J}\xi_J + i V_m^{(r)} \xi_I \;,
\end{equation}
and similarly for $\tilde \xi$.\footnote{Note that we set the $U(1)_r$ charge of $\xi_{I\alpha}$ to be $r=-1$, while for $\tilde\xi_I^{\dot\alpha}$ it is $r=+1$.}  These equations can be obtained by setting to zero the supergravity variations of the gravitini. Additionally, one should set to zero the variations of other fermionic fields in the supergravity multiplet. The resulting auxiliary equations, again omitting supergravity fields that are not turned on in our background,
\begin{equation}\label{auxKSE}
\sigma^m\tilde\sigma^n D_mD_n\xi_I = M\xi_I\;, \qquad \tilde\sigma^m\sigma^n D_mD_n\tilde\xi_I = M\tilde\xi_I
\end{equation}
are automatically satisfied.

The four solutions generate an $\mathfrak{su}(1|1)\oplus \mathfrak{su}(1|1)$ superalgebra. The appearance of this algebra does not come as a surprise: it is the Weyl transformation to $S^3\times \mathbb R$ of the flat-space $\mathfrak{su}(1|1)\oplus \mathfrak{su}(1|1)$-algebra mentioned in subsection \ref{subsection:schurlimitdefn}, which is preserved by compactifying $\mathbb R$ as in our background.

Let us describe the solutions explicitly. To that purpose, we first define the spinors $\kappa_{st}$, for $s=\pm 1, t=\pm 1$,
\begin{equation}
\kappa_{st}=\frac{1}{2}\begin{pmatrix}
e^{\frac{i}{2}(s\chi+t\varphi-st\theta)}\\
-s e^{\frac{i}{2}(s\chi+t\varphi+st\theta)}
\end{pmatrix}\;.
\end{equation}
In terms of these, a generic supercharge in $\mathfrak{su}(1|1)\oplus \mathfrak{su}(1|1)$ is described by the four-parameter family 
\begin{equation}\label{genericQ}
\mathfrak Q_{c_1,\tilde c_1}^{c_2,\tilde c_2} \longleftrightarrow \left\{\begin{matrix} \xi_1 = c_1\, \kappa_{+-}\;,\\ \xi_2 = c_2\, \kappa_{++}\;, \\\tilde\xi_1 = \tilde c_2\, \kappa_{--}\;, \\ \tilde\xi_2 = \tilde c_1\, \kappa_{-+}\;.\end{matrix} \right.
\end{equation}
Note that $\mathfrak Q_{c_1,\tilde c_1}^{0,0}$ are the supercharges of one copy of $\mathfrak{su}(1|1)$ and $\mathfrak Q_{0,0}^{c_2,\tilde c_2}$ of the other one.

For localization purposes, we should select a particular supercharge $\mathcal Q$ within $\mathfrak{su}(1|1)\oplus \mathfrak{su}(1|1)$. A choice of supercharge that lies in either $\mathfrak{su}(1|1)$, would admit a further refinement of the index to include its full grading by the three fugacities $p,q,t$ \cite{Gadde:2011uv}. Equivalently, when performing a further deformation of the supergravity background that only preserves a single $\mathfrak{su}(1|1)$, the Killing spinor solution describing that supercharge is still a solution to the further deformed Killing spinor equations \eqref{KSE} and \eqref{auxKSE}. The resulting computation of the partition function would give rise to the standard expression of the superconformal index as a matrix integral.\footnote{The analogous localization computations for four-dimensional $\mathcal N=1$ theories was performed in \cite{Peelaers:2014ima} and for $\mathcal N=4$ theories in \cite{Nawata:2011un}.} In this paper, we are after a localization that is geared towards the Schur limit of the superconformal index and that makes manifest its interpretation as the torus partition function of the associated chiral algebra. Therefore, we pick a supercharge that does not lie in either $\mathfrak{su}(1|1)$ algebra and with the property that it geometrically singles out a torus. More in detail, we define the supercharge $\mathcal Q_{c_1, c_2}$
\begin{equation}\label{ourQ}
\mathcal Q_{c_1,c_2} \colonequals \mathfrak Q_{c_1,c_2}^{c_2,c_1}\;.
\end{equation}
Any choice of $\mathcal Q_{c_1,c_2}$ with $c_i\neq 0$ is a good one for our purposes. To avoid clutter, starting in the next section, we will simply choose $c_1=c_2=1$ and denote $\mathcal Q \colonequals \mathcal{Q}_{1, 1}$. Note that the supercharge $\mathcal Q$ essentially corresponds to the flat space supercharge $\qq_1+\qq_2$. Let us finally define some useful bilinears of the Killing spinors which we will use throughout the paper
\begin{align}
&s \colonequals (\xi^I \xi_I)\;, \qquad \tilde s \colonequals (\tilde \xi_I\tilde \xi^I)\;, \qquad s' \colonequals (\xi'^I \xi'_I), \qquad \tilde s' \colonequals (\tilde \xi'_I \tilde \xi'^I), \qquad R^\mu \colonequals (\xi^I \sigma^\mu \tilde \xi_I)\;, \nn\\
&R^\mu_{IJ} \colonequals (\xi_I \sigma^\mu \tilde \xi_J)\;, \qquad (\Theta_{IJ})_{\mu \nu} \colonequals (\xi_I \sigma_{\mu \nu}\xi_J) \;,\qquad (\tilde \Theta_{IJ})_{\mu \nu} \colonequals (\tilde \xi_I \tilde \sigma_{\mu \nu}\tilde \xi_J) \;.\label{bilinears}
\end{align}
In particular, for the Killing spinors describing $\mathcal Q_{c_1,c_2}$, we have
\begin{equation}
s = - i c_1 c_2\sin\vartheta e^{i \chi}\;, \qquad
\tilde s = - i c_1 c_2\sin\vartheta e^{ - i \chi}\;, \qquad
R^\mu \partial_\mu = \frac{i c_1 c_2}{\ell} \partial_\chi\;.
\end{equation}
Note that the point-wise fixed locus of the vector field $R^\mu \partial_\mu$ is precisely the torus $T^2_{\vartheta = 0}$ when $c_i\neq 0$.

\subsection{Actions and supersymmetry transformation rules}
Let us briefly recall the Lagrangians and supersymmetry variation rules for the vector and hypermultiplet, see \cite{Hama:2012bg}. These expressions are implicitly evaluated on the supergravity background on $S^3\times S^1$ described in subsection \ref{subsection:sugra_background} (and we have omitted terms coupling to supergravity fields that are not turned on in our background). The Lagrangian describing the dynamics of an $\mathcal N=2$ vector multiplet is given by
\begin{multline}\label{VM-Lagr}
\mathcal{L}_{\text{YM}} =\tr{} \bigg[ \frac{1}{2}F_{\mu\nu}F^{\mu\nu} - 4 D_\mu \tilde{\phi} D^\mu \phi +  2M \phi \tilde{\phi} + 4 [\phi, \tilde{\phi}]^2 - \frac{1}{2} D_{IJ}D^{IJ}  \\
      - 2 i (\lambda^I \sigma^\mu D_\mu \tilde{\lambda}_I) - 2 \lambda^I [\tilde{\phi}, \lambda_I] + 2 \tilde{\lambda}^I [\phi, \tilde{\lambda}_I]
  \bigg] \;.
\end{multline}
It is invariant (up to total derivative terms) under the variations,
\begin{equation}\label{VM-SUSY}
\begin{aligned}
  \mathcal{Q}{A_\mu } = &\; + i({\xi ^I}{\sigma _\mu }{{\tilde \lambda }_I}) - i({{\tilde \xi }^I}{{\tilde \sigma }_\mu }{\lambda _I})  \\
  Q\phi  = &\;  - i({\xi ^I}{\lambda _I})  \\
  Q\tilde \phi  = &\;  + i({{\tilde \xi }^I}{{\tilde \lambda }_I})  \\
  \mathcal{Q}{\lambda _I} = &\; + \frac{1}{2}{F_{\mu \nu }}{\sigma ^{\mu \nu }}{\xi _I} + 2{D_\mu }\phi {\sigma ^\mu }{{\tilde \xi }_I} + \phi {\sigma ^\mu }{D_\mu }{{\tilde \xi }_I} + 2i{\xi _I}[\phi ,\tilde \phi ] + {D_{IJ}}{\xi ^J} \\
  \mathcal{Q}{{\tilde \lambda }_I} = &\; + \frac{1}{2}{F_{\mu \nu }}{{\tilde \sigma }^{\mu \nu }}{{\tilde \xi }_I} + 2{D_\mu }\tilde \phi {{\tilde \sigma }^\mu }{\xi _I} + \tilde \phi {{\tilde \sigma }^\mu }{D_\mu }{\xi _I} - 2i{{\tilde \xi }_I}[\phi ,\tilde \phi ] + {D_{IJ}}{{\tilde \xi }^J} \\
  \mathcal{Q}{D_{IJ}} = &\;  - i({{\tilde \xi }_I}{{\tilde \sigma }^\mu }{D_\mu }{\lambda _J}) + i({\xi _I}{\sigma ^\mu }{D_\mu }{{\tilde \lambda }_J}) - 2[\phi ,({{\tilde \xi }_I}{\lambda _J})] + 2[\tilde \phi ,({\xi _I}{\lambda _J})] + (I \leftrightarrow J)\;,
  \end{aligned}
\end{equation} 
In fact, as we show in more detail in appendix \ref{app:Q-exactness}, the Yang-Mills action is exact with respect to all supercharges in $\mathfrak{su}(1|1)\oplus\mathfrak{su}(1|1)$:
\begin{align}
S_{\text{YM}} &= \frac{1}{g_{\text{YM}}^2}\int_{S^3\times S^1} d^4x \sqrt{g}\ \mathcal{L}_{\text{YM}} + \frac{i\theta}{8\pi^2}\int_{S^3\times S^1} \tr{} F\wedge F \nn \\\label{YMQexact}
&= \mathfrak Q_{c_1,\tilde c_1}^{c_2, \tilde c_2}\Bigg[   (4\pi i) \int_{S^3\times S^1} d^4x \sqrt{g}\ (\tau_{\text{YM}} \Xi^- - \bar\tau_{\text{YM}} \Xi^+)  \Bigg]\;,
\end{align}
for all $c_i, \tilde c_i$. Here $\tau_{\text{YM}} = \frac{\theta}{2\pi} + \frac{4\pi i}{g_{\text{YM}}^2}$ is the standard exactly marginal coupling of the theory.\footnote{To be precise, these are the $\mathcal N=2$ supersymmetry preserving exactly marginal couplings, and there is one for each simple factor in the gauge group.} The precise form of $\Xi^\pm$ can be found in appendix \ref{app:Q-exactness}. An immediate consequence of this observation is that the partition function is independent of exactly marginal couplings, as it behooves an index. Similarly, observables preserved by a supercharge in $\mathfrak{su}(1|1)\oplus\mathfrak{su}(1|1)$ are necessarily also independent of couplings.

A collection of $N$ hypermultiplets, chosen such that the theory is superconformal, coupled to a dynamical or background vector multiplet has Lagrangian
\begin{multline}\label{HM-Lagr}
\mathcal{L}_{\text{HM}}=\bigg[ \frac{1}{2}{D_\mu }{q^{IA}}{D^\mu }{q_{IA}} - q^{IA} \{ \phi, \tilde \phi\}{_A}{^B} q_{IB} +\frac{i}{2}{q^{IA}}(D_{IJ})_A{^B}{q^J}_B + \frac{\mathcal R}{12}{q^{IA}}{q_{IA}} \\
  - \frac{i}{2}{{\tilde \psi }^A}{{\tilde \sigma }^\mu }{D_\mu }{\psi _A}  - \frac{1}{2}{\psi ^A}{\phi _A}^B{\psi _B} + \frac{1}{2}{\tilde \psi ^A}{\tilde \phi }{_A}^B{\tilde \psi _B}  - {q^{IA}}((\lambda _I)_A{^B}{\psi _B}) + ({\tilde \psi ^A} (\tilde\lambda ^I){_A}{^B}){q_{IB}}\bigg]\;,
\end{multline}
where $\mathcal R$ is the Ricci scalar, which evaluates to $\mathcal R = \frac{6}{\ell^2}$ on our background. The resulting action is invariant under the variations
\begin{equation}\label{HM-SUSY}
\begin{aligned}
\mathcal{Q} q_{IA} = &\ - i(\xi_I\psi _A) + i(\tilde \xi_I\tilde \psi_A) \\
\mathcal{Q}\psi_A = &\ + 2D_\mu q_A^I\sigma^\mu\tilde \xi_I - 4iq_A^I \xi'_I - 4 i \xi_I \tilde \phi_A{^B} q^I_B + 2\check \xi_{I'}F^{I'}_A \\
\mathcal{Q}\tilde \psi_A = &\ + 2D_\mu q_A^I \tilde \sigma^\mu \xi_I - 4iq_A^I \tilde \xi'_I - 4 i \tilde \xi_I \phi_A{^B} q^I_B  + 2\tilde{\check \xi} _{I'} F^{I'}_A  \\
\mathcal{Q}{F_{I'A}} = &\ + i (\check \xi_{I'}\sigma^\mu D_\mu \tilde \psi_A) - i( \tilde{\check \xi}_{I'} \tilde \sigma^\mu D_\mu \psi _A )\\
    & \ - 2 \phi_A^B(\check \xi_{I'} \psi_B) - 2 (\check \xi_{I'} \lambda_J)_A^B q^J_B + 2 \tilde \phi_A^B (\tilde {\check \xi}_{I'} \tilde \psi_B) + 2 (\tilde {\check \xi}_{I'} \tilde\lambda_J)_A^B q^J_B \;,
  \end{aligned}
\end{equation}
Here, the indices $A,B,\ldots$ on the hypermultiplet fields denote $USp(2N)$ indices; the gauge and/or flavor groups are embedded in this $USp(2N)$. The supersymmetry variations also contain the spinors ${\check \xi}_{I'}$ and $\tilde{\check \xi}_{I'}$, where the primed indices $I',J',\ldots$ are acted on by another $SU(2)_{R'}$ group. The introduction of these checked spinors is necessary to achieve an off-shell description of the hypermultiplet. They must satisfy the constraints
\begin{equation}\label{constraintsoncheckedspinors}
  (\xi_I\check\xi_{J'})-(\tilde \xi_I\tilde{\check\xi}_{J'})=0\;, \;\; (\xi^I\xi_I)+(\tilde{\check\xi}^{I'}\tilde{\check\xi}_{I'})=0\;, \;\; (\tilde\xi^I\tilde\xi_I)+( {\check\xi}^{I'}{\check\xi}_{I'})=0\;, \;\; (\xi^I \sigma^\mu \tilde\xi_I)+({\check\xi}^{I'}\sigma^\mu\tilde{\check\xi}_{I'})=0\;.
\end{equation}
Note that these equations do not have a unique solution for ${\check \xi}_{I'}$ and $\tilde{\check \xi}_{I'}$, as reflected in their covariance under $SU(2)_{R'}$ rotations acting on the primed indices. We will make a particular choice for the checked spinors, locking the $SU(2)_R$ and $SU(2)_{R'}$ rotations, namely
\begin{equation}
  \check\xi_I' = - e^{-i\chi} \xi_I\;, \qquad \tilde{\check{\xi}}_{I'} = e^{ + i \chi} \tilde\xi_I \ .
\end{equation}
The supercharge described in the previous subsection, see \eqref{ourQ}, squares to
\begin{equation}\label{ourQsquared}
\mathcal Q_{c_1,c_2}^2 = -2i \mathcal{L}_R^{A + V} + \text{Gauge}(\Phi) + \text{R}_r(\Theta)\;,
\end{equation}
Here $\mathcal{L}_R^{A + V}$ is a gauge covariant Lie derivative along the vector field $R=\frac{i c_1 c_2}{\ell}\partial_\chi$. It is covariant with respect to dynamical gauge fields and background flavor fields, and the background $SU(2)_\mathcal{R}$, $U(1)_r$ and $SU(2)_{\check{\mathcal{R}}}$ gauge fields. The parameter of the additional gauge transformation reads $\Phi = 2i(s \tilde \phi + \tilde s \phi)$. Finally, the $U(1)_r$ rotation has parameter $\Theta = - (\xi^I \xi'_I) + (\tilde \xi_I \tilde \xi'^I) = - \frac{i c_1 c_2}{\ell}$.

It will turn out to be useful to rewrite the terms in the action involving auxiliary fields in terms of a new set of variables. Let us make some invertible redefinitions,
\begin{equation}\label{def:Dz}
\mathfrak{D}_z \colonequals \frac{\ell}{2i}D_{IJ}(R^{IJ}_4 + i R^{IJ}_1)\;, \quad \mathfrak{D}_{\bar z} \colonequals \frac{2i}{\ell}\frac{1}{s \tilde s}D_{IJ}(R^{IJ}_4 - i R^{IJ}_1)\;, \quad \mathfrak{D}_3 \colonequals \frac{1}{\sqrt{s \tilde s}} D_{IJ}R^{IJ}_3\;,
\end{equation}
and similarly we define $(\mu_{z,\bar z, 3})_A^{\phantom{A}B}$ by replacing $D_{IJ}$ in the above definitions with the moment map operator $(\mu_{IJ})_A^{\phantom{A}B}=q_{(IA}q_{J)}{^B}$. Using a Fierz identity (see \eqref{fierz}), we rewrite the terms with auxiliary fields as follows
\begin{align}\label{Dcontents}
\mathcal{L}_\text{HM} + \mathfrak{s} \mathcal{L}_\text{YM}\Big|_{D} = & \tr\bigg[- \mathfrak{s}\mathfrak{D}_z \mathfrak{D}_{\bar z} - \mathfrak{s}\mathfrak{D}_3\mathfrak{D}_3 + \frac{i}{2}\mu_z\mathfrak{D}_{\bar z} + \frac{i}{2}\mu_{\bar z}\mathfrak{D}_{z} + i \mu_3 \mathfrak{D}_3 \bigg] \\
  = & \tr\bigg[
    - \mathfrak{s} \big(\mathfrak{D}_z - \frac{i}{2\mathfrak{s}}\mu_z\big)\big(\mathfrak{D}_{\bar z} - \frac{i}{2\mathfrak{s}}\mu_{\bar z}\big)
    - \mathfrak{s} \big(\mathfrak{D}_3 - \frac{i}{2\mathfrak{s}}\mu_3\big)^2
    - \frac{1}{4\mathfrak{s}^2}(\mu_z\mu_{\bar z} + \mu_3\mu_3)\bigg]\;.\nn
\end{align}
For future purposes, we have included an undetermined relative coefficient $\mathfrak{s}$ between the hypermultiplet Lagrangian and the Yang-Mills Lagrangian.

\section{Localizing the Schur index}\label{section:localizingSchur}
In this section we perform a supersymmetric localization computation on the supergravity background defined in the previous section with the aim of finding a path-integral derivation of the interpretation of the Schur index as a (graded) torus partition function. To that purpose, we choose as localizing supercharge $\mathcal Q$ the one identified in \eqref{ourQ}.

Let us start by briefly recapping the localization argument \cite{Witten:1988ze,Witten:1991zz}.\footnote{See also the review \cite{Pestun:2016zxk} and its contributions.} The Euclidean path-integral over the space of fields can be deformed by a $\mathcal Q$-exact deformation
\begin{equation}\label{keyLocStep}
\int [D\Phi] e^{-S[\Phi]} = \int [D\Phi] e^{-S[\Phi]-\mathfrak s\, \mathcal{Q}\int V}\;,
\end{equation}
if $\mathcal{Q}$ is a symmetry of the action, $\mathcal{Q}S=0$, and also $\mathcal{Q}^2 \int V=0$. Here $V$ is any fermionic functional. The reasoning is standard:\footnote{Note that implicitly we also assume that the path integral measure is invariant under supersymmetry.}
\begin{equation}
\frac{d}{d\mathfrak s} \int [D\Phi] e^{-S[\Phi]-\mathfrak s\, \mathcal{Q}\int V} = -\int [D\Phi]\, \mathcal{Q}\big(({\textstyle \int V})\, e^{-S[\Phi]-\mathfrak s\, \mathcal{Q}\int V} \big) = 0\;.
\end{equation}
If the real part of the bosonic piece of the deformation Lagrangian is positive definite, $\re \mathcal{Q}V\big|_{\text{bos.}}\geq 0$, the path integral will localize to the zeros of $\re \mathcal{Q}V\big|_{\text{bos.}}$. A one-loop approximation around this locus is exact, as can be argued by sending the parameter $\mathfrak s$ to infinity. Denoting the set of (bosonic) field configurations solving $\re \mathcal{Q}V\big|_{\text{bos.}}= 0$ as $\{ \phi_0 \}$, the result of the localization computation can thus be written as 
\begin{equation}
\int [D\Phi] e^{-S[\Phi]} = \SumInt_{\{\phi_0\}} e^{-S[\phi_0]}\; Z_{\text{1-loop}}[\phi_0]\;.
\end{equation}
Our task is thus, in order, to find the localization locus $\{ \phi_0 \}$, and to evaluate the classical action and one-loop determinant of quadratic fluctuations. In section \ref{section:insertions}, we will elaborate in great detail on the compatibility of the localization computation with the insertion of (local) operators in the path integral and explain how the result should be modified to incorporate such insertions.

\subsection{Localization locus}
Our first task is to determine the localization locus. We take the canonical choice for $V$, \ie{},\footnote{This choice of $V$ is associated to so-called Coulomb branch localization computations. It would be interesting to analyze different choices of $V$. A particularly interesting alternative choice that has been explored in detail in the literature, gives rise to so-called Higgs branch localization computations \cite{Benini:2012ui,Doroud:2012xw,Fujitsuka:2013fga,Benini:2013yva,Peelaers:2014ima,Pan:2014bwa,Chen:2015fta,Pan:2015hza}.}
\begin{equation}\label{standarddefferm}
V = \sum_{\text{fermions }\psi} (\mathcal Q \psi)^\dagger \, \psi\;.
\end{equation}
Consequently, the localization locus is determined by the BPS equations $\mathcal Q \psi=0$, for all fermions $\psi$, supplemented by the reality conditions that define $\dagger$. These equations are easy to solve. 

We impose the following reality conditions on the bosonic fields in the vector multiplet
\begin{equation}
\phi ^\dagger =  - \tilde \phi\;, \qquad A_\mu^\dagger  = A_\mu\;, \qquad D_{IJ}^\dagger = - D^{IJ} \label{reality-condition-phi-F}\;.
\end{equation}
The BPS equations $\mathcal Q \lambda_I=\mathcal Q \tilde \lambda_I=0$ in \eqref{VM-SUSY} can then be split into real and imaginary parts and, together with the Bianchi identity, only allow flat connections as globally smooth solutions,\footnote{There are in principle non-smooth solutions with more nontrivial profiles}
\begin{align}\label{locusVM}
F_{\mu\nu} = 0\;, \qquad \phi = \tilde\phi = D_{IJ} = 0\;.
\end{align}
The topology of $S^3 \times S^1$ allows nontrivial solutions to $F_{\mu\nu} = 0$ given by $A_\mu dx^\mu = \mathfrak a dt$ with $\mathfrak a$ a constant in the Cartan subalgebra, up to constant gauge transformations. Such flat connections were already used in describing the flavor symmetry background, see \eqref{flavorbg}. Note that the Yang-Mills action \eqref{VM-Lagr} itself can be used as localizing action for the vector multiplet as it is both $\mathcal{Q}$-exact and the real part of its bosonic terms is positive semi-definite under the reality properties \eqref{reality-condition-phi-F}.\footnote{Recall that we chose the parameters $\beta_1$ and $\beta_2$ to be purely imaginary such that the $U(1)_r$ connection is real.} One easily confirms the localization locus this way.

We subject the bosonic fields in the hypermultiplet to the reality conditions $q_{IA}^\dagger = \epsilon^{IJ}\Omega^{AB} q_{JB}$ and $F_{I'A}^\dagger = - \epsilon^{I'J'} \Omega^{AB} F_{J'B}$, where $\Omega$ is a symplectic form. Setting the vector multiplet fields to their BPS values (\ref{locusVM}), one can reorganize the BPS equations $\mathcal Q \psi_A = \mathcal Q \tilde\psi_A=0$, see \eqref{HM-SUSY}, into a set of partial differential equations for $q_{IA}$, and a set of algebraic equations that solve $F_{I'A}$ in terms of $q_{IA}$. In particular, one can show that
\begin{equation}\label{chi-independence}
R^\mu D_\mu q_{IA} = R^\mu D_\mu F_{I'A} = 0 \quad \Longleftrightarrow  \quad D_\chi q_{IA} =  D_\chi F_{I'A} = 0 \;.
\end{equation}
The remaining first-order partial differential equations for $q_{IA}$ lead to one second-order elliptic partial differential equation for each separate component $q_{IA}$, ensuring that its profile is completely specified by its value on the torus $T^2_{\vartheta=0}$. In other words, the localization locus is given by the space of complex functions on the torus capturing the boundary profiles of half the components of $q_{1A}$. The remaining components of $q$ are determined from these either by complex conjugation or through the (first-order) BPS equations.

The result of the localization computation will be an integral over the localization locus described in this subsection. Its integrand contains the classical action evaluated on the localization locus and the one-loop determinants of quadratic fluctuations around the localization locus. We will address these two computations in the next subsection.

\subsection{Evaluation of the classical action and one-loop determinant}
The evaluation of the Yang-Mills action on the localization locus \eqref{locusVM} is trivial,
\begin{equation}
S_{\text{YM}}\big|_{\text{loc. locus}} = 0\;.
\end{equation}
The one-loop determinant of quadratic fluctuations of the vector multiplet fields can be copied from the literature. The reason is that the Yang-Mills action can be used as localizing action with respect to all supercharges in $\mathfrak{su}(1|1)\oplus \mathfrak{su}(1|1)$, as it is exact and the real part of its bosonic terms is positive semi-definite. The localization locus and vector multiplet one-loop determinant are thus shared among all choices of localizing supercharge, while the classical action trivially evaluates to zero on the locus. Note however that different choices of localization supercharge can allow for different insertions. We will analyze such insertions in more detail in section \ref{section:insertions} and explain how the computation can accommodate them. We thus have\footnote{Note that this one-loop determinant also includes contributions from the gauge-fixing ghosts.}
\begin{equation}\label{oneloopVM}
Z_{\text{1-loop}}^{\text{VM}}(\mathfrak b,\tau) = (-i)^{\rank \mathfrak g - \dim \mathfrak g}\ \eta(\tau)^{-\dim \mathfrak g + 3 \rank \mathfrak g}\prod_{\alpha\neq 0} \theta_1(\alpha(\mathfrak{b})|\tau)\;,
\end{equation}
where the product runs over the nonzero roots of the Lie-algebra $\mathfrak g$ associated with the gauge group, and we expressed the result in terms of the Dedekind eta function $\eta(\tau)$ and one of the Jacobi theta functions.\footnote{See appendix \ref{specialfuctions} for definitions and useful properties of these functions.}

The evaluation of the hypermultiplet action on the localization locus is nontrivial, but as the computation is mostly the same as the one performed in \cite{Pan:2017zie}, we will be brief. As was observed in for example \cite{Dedushenko:2016jxl,Bonetti:2016nma}, it is sufficient to substitute in the action \eqref{HM-Lagr} the solutions of the complex BPS equations for the auxiliary fields $F_{I'A}$ and $D_{IJ}$, which come in two versions each,
\begin{align}
F_{I'A} = - \frac{1}{\tilde s}\Big[ 2(\check \xi_{I'} \sigma^\mu \tilde \xi_J) D_\mu q^J{_A} + (\check \xi_{I'} \sigma^\mu D_\mu \tilde\xi_J) q^J{_A}  + 4 i (\check \xi_{I'} \xi_J) \tilde \phi^A{_B}q^{JB}  \Big] \;,\\
F_{I'A} = + \frac{1}{s}\Big[ 2(\tilde{\check \xi}_{I'} \sigma^\mu \xi_J) D_\mu q^J{_A} + (\tilde{\check \xi}_{I'} \tilde\sigma^\mu D_\mu \xi_J) q^J{_A} + 4 i (\tilde{\check \xi}_{I'} \tilde \xi_J) \phi^A{_B}q^{JB} \Big] \;,
\end{align}
and
\begin{align}
D_{IJ} = - \frac{1}{2s^2}\Big[-2s F_{\mu \nu} + 4(R \wedge d_A\phi)_{\mu\nu} - 4i (\xi^I \sigma_{\mu \nu} \xi'_I) \phi \Big] \Theta^{\mu \nu}_{IJ} \; ,\\
  D_{IJ} = +\frac{1}{2 \tilde s^2}\Big[  -2 \tilde s F_{\mu \nu} + 4(R \wedge d_A \tilde \phi)_{\mu\nu} + 4i (\tilde \xi^I \tilde \sigma_{\mu \nu} \tilde \xi'_I) \tilde\phi \Big] \tilde \Theta^{\mu \nu}_{IJ} \; .
\end{align}
Here we used various bilinears of the Killing spinors as defined in \eqref{bilinears}. By a slight abuse of notation, we introduced the one-form $R=R_\mu dx^\mu$. Also using that the Killing spinors describing our supercharge \eqref{ourQ} satisfy the Killing spinor equations and additionally that the relations
\begin{equation}
(\xi_I\xi'_J) = (\tilde\xi_I \tilde \xi'_J)\;, \qquad d R = 0 \; , \qquad \frac{2 \tilde s'}{s} + \frac{2}{s \tilde s}R^\mu (\xi'^I \sigma_\mu \tilde \xi'_I) = \frac{1}{8} \mathcal{R} + \frac{3M}{8} \;,
\end{equation}
hold on our background, it can be shown by straightforward computation that on the complex localization locus the hypermultiplet action evaluates to
\begin{multline}\label{HMactiononLL}
  S = \int_{S^3 \times S^1}d^4x\, \sqrt{g}\,  D_\mu \bigg[  \frac{1}{s\tilde s} \epsilon^{\mu \nu \lambda \delta} R_\lambda \big( (q^I{_A}\xi_I) \sigma_\delta D_\nu  (q^{KA}  \tilde \xi_K) \big) - \frac{2i}{s\tilde s} R^\mu (q^{KA}\xi'_K)(q^I{_A}\xi_I) \\+ \frac{i}{s \tilde s} (s \tilde\phi - \tilde s\phi)_{AB}R^\mu_{IJ} q^{IA}q^{JB}\bigg]\;.
\end{multline}
The quantity in the square bracket is independent of the coordinate $\chi$. This follows immediately from the observation that it is gauge invariant and, if we also transform the Killing spinors according to their assigned charges, neutral under $U(1)_r$ and $SU(2)_R$. Indeed, on the complex localization locus, all fields are invariant under $\mathcal Q^2$, see \eqref{ourQsquared}, and so are the Killing spinors. Then it is clear that neutral, gauge invariant objects are necessarily independent of $\chi$. This has the immediate consequence that the $D_\chi$ derivative gives zero and that we can trivially perform the integral over the coordinate $\chi$. At this point, one is left with an integral of a three-dimensional total derivative over a subspace which is topologically $D^2 \times S^1$, parametrized by the coordinates $\varphi, \vartheta$ and $t$.\footnote{We choose this subspace to be specified by $\chi = 0$.} Its boundary is precisely the torus $T^2_{\vartheta=0}$ parametrized by $\varphi$ and $t$. Applying Stokes' theorem results in
\begin{equation}\label{HM2daction}
S_\text{SB}^{T^2}[q] = 2\pi i \ell\int_{T^2} d^2 x \sqrt{\det \widehat{g_{T^2}}} \Big[\epsilon^{\mu \nu}(q^I{_A}\xi_I) \sigma_\mu D_\nu (q^{JA}\tilde \xi_J) \Big]\ .
\end{equation}
Note that the components of the epsilon tensor evaluate to $\epsilon^{\varphi t} = -\epsilon^{t\varphi} =(\det \widehat{g_{T^2}})^{-1/2}.$ This action suggests that the bosonic spinors $\mathfrak{q}_A\equiv q^I{_A} \xi_I$ restricted to live on the torus $T^2$ should be treated as the fundamental field variables.\footnote{Similar to the analysis in \cite{Pan:2017zie}, the four-dimensional spinors $q^I_A \xi_I$ and $q^I{_A} \tilde \xi_I$ are proportional to each other when restricted to the torus $T^2$, hence only one set of independent field variables $\mathfrak{q}_A$ is needed. Also note that the localization locus relates $\mathfrak{q}_1$ and $\mathfrak{q}_2$ via a (nontrivial) reality property.} By construction, the fields $\mathfrak{q}_A$ are periodic in $t$ but anti-periodic in $\varphi$, precisely exhibiting their spinorial nature. Also observe that while this action naively still depends on the parameters $\beta_1$ and $\beta_2$ through the $U(1)_r$ connection in the covariant derivative, the connection piece will drop out in the antisymmetrization imposed by the epsilon-tensor. The $\beta$-dependence of the measure cancels trivially against the hidden metric dependence of the curved epsilon tensors, leaving us with a manifestly $\beta$-independent action. Having achieved independence of the $\beta$s, we can tune them without loss of generality to the K\"ahler limit discussed below \eqref{induced-torus-metric}.

For computational purposes, it is often more convenient to expand \eqref{HM2daction} in its spinorial components. Let us thus define\footnote{For convenience, we absorbed some additional factors in the definitions of $Q$ and $\widetilde Q$.}
\begin{equation}\label{defnQQtilde}
Q \colonequals 2\pi e^{-\frac{\pi i}{4}}\sqrt{\ell}\ \big( e^{\frac{i\varphi}{2}} q_{11} - e^{-\frac{ i\varphi}{2}} q_{21}\big)\;, \qquad  
  \widetilde Q \colonequals 2\pi e^{-\frac{\pi i}{4}}\sqrt{\ell}\ \big(e^{\frac{i\varphi}{2}} q_{12} -  e^{- \frac{i\varphi}{2}} q_{22}\big) \;.
\end{equation}
and write the action explicitly as
\begin{equation}\label{HMactionSB}
S_\text{SB}^{T^2}[Q, \widetilde Q] = -\frac{1}{2\pi\ell} \int  \sqrt{g}d^2z \Big[ Q\,  D_{\bar z}^{\mathfrak A}\widetilde Q - \widetilde Q\, D_{\bar z}^{\mathfrak A} Q\Big] =  \frac{1}{\pi \ell} \int  \sqrt{g}d^2z  \, \widetilde Q\, D_{\bar z}^{\mathfrak A} Q  \; .
\end{equation}
This action is defined on a torus with complex structure $\tau$, and parametrized by the complex coordinates $z = \varphi + \tau t$, $\bar z = \varphi + \bar \tau t$. The fields $Q$ and $\widetilde Q$ respectively transform in a representation $\mathfrak R$ and $\mathfrak R^*$ of the combined gauge and flavor group with associated flat connection $\mathfrak A$. The flavor and gauge potentials $\mathfrak{a}$ and $\mathfrak{b}$ are thus collectively denoted by $\mathfrak{A}$. Note that the differential operator in the action governing the dynamics of $Q$ and $\widetilde Q$ is simply the antiholomorphic covariant derivative\footnote{Similarly, we have for the holomorphic covariant derivative $D_z = \partial_z - i \frac{A_t - \tilde \tau A_\varphi}{2i\im \tau} = \frac{1}{2i\im \tau}(\partial_t - \bar \tau \partial_\varphi)- i \frac{A_t - \tilde \tau A_\varphi}{2i\im \tau}$ and $D_z^{\mathfrak A} = \partial_z -i \frac{\mathfrak{A}}{2i\im\tau}$. Note that our derivatives are dimensionless. To make it into a more standard dimension one object, we should multiply it with a factor $\ell^{-1}$.}
\begin{equation}
D_{\bar z} = \partial_{\bar z} -i \frac{A_t - \tau A_\varphi}{\bar \tau-\tau} = - \frac{1}{2i\im \tau}(\partial_t - \tau \partial_\varphi) +i \frac{A_t - \tau A_\varphi}{2i\im \tau}\;,
\end{equation}
evaluated on the BPS locus $A=\mathfrak A dt$. To avoid clutter, we introduced some notation for these derivatives evaluated on the locus
\begin{equation}
D_{\bar z}^{\mathfrak A} = \partial_{\bar z} +i \frac{\mathfrak{A}}{2i \im\tau} \;.
\end{equation}

A direct computation of the one-loop determinant of quadratic fluctuations for the hypermultiplet around the locus described above is technically challenging. Fortunately, one can argue indirectly that the answer is simply equal to 1,
\begin{equation}\label{HMoneloop}
Z_{\text{1-loop}}^{\text{HM}}(\mathfrak A,\tau) = 1\;.
\end{equation}
Before presenting the argument in the next subsection, we collect all pieces and present the final result of the localization computation of the partition function (in the absence of any insertions, which will be analyzed in detail in section \ref{section:insertions})
\begin{equation}\label{finalresult}
Z_{S^3\times S^1}(\mathfrak a,\tau) = \frac{1}{|W|} \oint \prod_{j=1}^{\rank \mathfrak g} \frac{db_j}{2\pi i b_j}\ Z_{\text{1-loop}}^{\text{VM}}(\mathfrak b,\tau) \int[DQ D\widetilde Q] \ e^{- S^{T^2}_\text{SB}[Q, \widetilde Q]}\;,
\end{equation}
where $|W|$ denotes the order of the Weyl group of $\mathfrak g$. Further, $b_j = \exp[2\pi i \mathfrak b_j]$ are the $\rank \mathfrak g$ gauge fugacities. These are integrated over along the unit circle. As far as the hypermultiplet sector is concerned, we have manifestly recovered a symplectic boson pair residing on the torus $T^2$.

\subsection{Hypermultiplet one-loop determinant}\label{HMoneloopcomputation}
In the absence of any insertions, the result \eqref{finalresult} for the partition function should agree with the standard computation of this quantity as a matrix model. For a gauge theory with gauge group $G$ and hypermultiplets transforming in $\rank \mathfrak{g}_F$ irreducible representations $\mathcal R_l$ of the gauge group, it reads
\begin{align}\label{finalresultOLD}
Z_{S^3\times S^1}(\mathfrak a,\tau) = \frac{1}{|W|} \oint \prod_{j=1}^{\rank \mathfrak g} \frac{db_j}{2\pi i b_j}\ &(-i)^{\rank \mathfrak g - \dim \mathfrak g}\ \eta(\tau)^{-\dim \mathfrak g + 3 \rank \mathfrak g}\prod_{\alpha\neq 0} \theta_1(\alpha(\mathfrak{b})|\tau) \nn\\
&\times \prod_{l=1}^{\rank \mathfrak{g}_F} \prod_{\rho\in \mathcal{R}_l}\frac{\eta(\tau)}{\theta_4(\rho(\mathfrak b)+\mathfrak a_l|\tau)}\;. 
\end{align}
Here $\rho\in \mathcal R_l$ are the weights of the representation $\mathcal R_l$. The quantity on the second line is the one-loop determinant of quadratic fluctuations of the hypermultiplet as obtained in the standard localization computation, \ie{}, the computation with respect to the supercharge $\mathfrak Q_{c_1,\tilde c_1}^{0,0}$ or $\mathfrak Q_{0,0}^{c_2,\tilde c_2}$, around the standard localization locus which sets all hypermultiplet fields to zero. Demanding the two expressions \eqref{finalresult} and \eqref{finalresultOLD} for the partition function to match allows to verify the absence of a one-loop determinant for the hypermultiplet in our current localization computation. In other words, equation \eqref{HMoneloop} is correct if
\begin{equation}\label{HMcheck}
\prod_{\rho\in \mathfrak{R}}\frac{\eta(\tau)}{\theta_4(\rho(\mathfrak A)|\tau)} =  \int[DQ D\widetilde Q] \ e^{- S^{T^2}_\text{SB}[Q, \widetilde Q]} \equalscolon Z^{\text{HM}}(\mathfrak A,\tau)\;,
\end{equation}
where $\mathfrak R = \bigoplus_{l=1}^{\rank \mathfrak{g}_F} \mathcal R_l$ and as before $\mathfrak A$ collectively denotes the flavor and gauge potentials $\mathfrak{a}$ and $\mathfrak{b}$. Let us thus compute the right-hand side, which we have called $Z^{\text{HM}}(\mathfrak A,\tau)$. Note that this quantity can be interpreted as the partition function of a collection of $\dim \mathfrak{R}$ free hypermultiplets.

We could evaluate $Z^{\text{HM}}(\mathfrak A,\tau)$ by simply performing the Gaussian integral. The result takes the intermediate form of an infinite product of eigenvalues, which should be regularized carefully. Instead, we take a more elegant route by performing a variation of the path integral with respect to the parameters $\tau$ and $\mathfrak A$ and evaluating the resulting integrated two-dimensional stress tensor and affine flavor current one-point functions explicitly. 

Let us start by taking the $\tau$-derivative of $Z^{\text{HM}}(\mathfrak A,\tau)$,
\begin{equation}
\frac{\partial}{\partial \tau} Z^{\text{HM}}(\mathfrak A,\tau) =  \frac{1}{\pi(\tau - \bar \tau)} \int[DQ D\widetilde Q] \left(\int_{T^2_{\tau}}  \sqrt{g}\ d^2z \ T_{\mathfrak A}(z)\right) \ e^{-\frac{1}{\pi\ell} \int_{T^2_{\tau}} \sqrt{g}\ d^2z\ \widetilde Q \left(\partial_{\bar z} - i \frac{\mathfrak A}{\bar\tau - \tau}\right)Q}\;,
\end{equation}
where the covariant holomorphic stress tensor $T_{\mathfrak A}$ is given by
\begin{equation}\label{covST}
T_{\mathfrak A}(z) \colonequals \frac{1}{2\ell}\left(Q \,D_z^{\mathfrak A} \widetilde Q - \widetilde Q\, D_z^{\mathfrak A} Q\right)(z)\;.
\end{equation}
To obtain this result, we used that $\frac{\partial}{\partial \tau}\partial_{\bar z} = \frac{1}{\tau-\bar \tau}\partial_z$. Using translational invariance, one obtains the first-order partial differential equation\footnote{The volume of the torus is easily computed:
\begin{equation}
\sqrt{g} \ dz \wedge d \bar z = + \ell^2 \im{\tau} \ d\varphi \wedge dt \qquad \Longrightarrow \qquad \text{Vol}(T^2) = \int_{T^2} \sqrt{g}\ dz \wedge d\bar z = (2\pi\ell)^2 \im{\tau} \;.
\end{equation}}
\begin{equation}\label{comparebefore}
  \frac{\partial}{\partial \tau} \ln Z^{\text{HM}}(\mathfrak A,\tau)
  =  \frac{1}{\pi(\tau - \bar \tau)}\vol(T^2_{\tau}) \left\langle T_{\mathfrak A}(0) \right\rangle_{T^2_\tau} = -2\pi i\ell^2 \left\langle T_{\mathfrak A}(0) \right\rangle_{T^2_\tau}\;.
\end{equation}
Our obvious next step is to evaluate the one-point function of the covariant stress tensor. To do so, we start by computing the flavored propagator 
\begin{equation}
(G_{\mathfrak A})_{\phantom{\rho}\rho}^{\rho'}(z-w,\bar z-\bar w|\tau) \colonequals \langle \widetilde Q^{\rho'}(z,\bar z) Q_\rho(w,\bar w) \rangle_{T^2_\tau}\;.
\end{equation}
Note that $G_{\mathfrak A}$ is a matrix in gauge/flavor indices and we decompose $Q$ and $\widetilde Q$ along the weights of the representation $\mathfrak{R}$ and its complex conjugate. The propagator $G_{\mathfrak A}$ should solve the differential equation\footnote{The Dirac $\delta$-function $\delta(z-w, \bar z-\bar w)$ on $T^2$ is characterized by
\begin{align}
  \int_{T^2} \sqrt{g}\ d^2w\ \delta(z-w, \bar z-\bar w) \ f(w, \bar w) = f(z,\bar z) \;.
\end{align}
}
\begin{equation}
\frac{1}{\pi\ell} \big(\partial_{\bar w} - i \frac{\rho(\mathfrak A)}{\bar\tau - \tau}\big) (G_{\mathfrak A})_{\phantom{\rho}\rho}^{\rho'}(z-w,\bar z-\bar w|\tau) = \delta(z-w,\bar z-\bar w) \; \delta_{\rho}^{\rho'}\;.
\end{equation}
It should also satisfy the appropriate double (anti)periodicity conditions
\begin{equation}
G_{\mathfrak A}(z + 2\pi (m + n \tau),\bar z + 2\pi (m + n \bar \tau)|\tau) = (-1)^m G_{\mathfrak A}(z,\bar z|\tau)\;,\qquad \text{for } m,n\in \mathbb Z\;.
\end{equation}
The solution is given by
\begin{equation}\label{propagator}
  (G_{\mathfrak A})_{\phantom{\rho}\rho}^{\rho'}(z,\bar z|\tau) =
 - \frac{1}{2\pi\ell}e^{- i\frac{\rho(\mathfrak{A})}{ \bar \tau- \tau} (\bar z - z)} \frac{\theta'_1(0|\tau)}{\theta_4(\rho(\mathfrak A)|\tau)} \frac{\theta_4(\frac{z}{2\pi} - \rho(\mathfrak A) | \tau)}{\theta_1(\frac{z}{2\pi}|\tau)}\; \delta_\rho^{\rho'}\; .
\end{equation}
Here the prime on the theta function denotes a derivative with respect to the first argument. The propagator $(G_{\mathfrak A})_{\rho}^{\phantom{\rho}\rho'}(z-w,\bar z-\bar w|\tau) \colonequals \langle Q_\rho(z,\bar z) \widetilde Q^{\rho'}(w,\bar w) \rangle_{T^2_\tau}$ is easily seen to be given by $(G_{\mathfrak A})_{\rho}^{\phantom{\rho}\rho'}(z-w,\bar z-\bar w|\tau) = -(G_{-\mathfrak A})_{\phantom{\rho}\rho}^{\rho'}(z-w,\bar z-\bar w|\tau) =  (G_{\mathfrak A})_{\phantom{\rho}\rho}^{\rho'}(w-z,\bar w-\bar z|\tau) $.

We proceed by covariantly point-splitting the covariant stress tensor \eqref{covST}. In other words, we connect the split insertions by a Wilson line to maintain the overall singlet transformation property under the (background) gauged symmetries. The standard Wilson line reads
\begin{equation}
\mathcal W_{\rho}^{\phantom{\rho}\rho'}[(0,0)\leftarrow (w,\bar w)] = \Big[\text{Pexp}\big(i\int_{(w,\bar w)}^{(0,0)} A\big) \Big]_{\rho}^{\phantom{\rho}\rho'}
\end{equation}
where the integral runs along a specific path connecting the points $(w,\bar w)$ and $(0,0)$, and we expressed the adjoint valued quantity as a matrix. Evaluated on the localization locus of constant $A$, it becomes
\begin{equation}
(\mathcal{W}_\mathfrak{A})_{\rho}^{\phantom{\rho}\rho'}[(0,0)\leftarrow (w,\bar w)] = e^{-i\frac{\rho(\mathfrak{A})}{\bar\tau-\tau} (\bar w - w)} \delta_{\rho}^{\rho'}
\end{equation}
which is independent of the chosen path between the two points as the field strength vanishes. Similarly, we find 
\begin{equation}
(\mathcal{W}_\mathfrak{A})_{\phantom{\rho}\rho}^{\rho'}[(0,0)\leftarrow (w,\bar w)] =e^{i\frac{\rho(\mathfrak{A})}{\bar\tau- \tau} (\bar w - w)} \delta_{\rho}^{\rho'}\;.
\end{equation}
The covariantly point-split stress tensor then becomes
\begin{align}
\langle T_{\mathfrak A}(0)\rangle \rightarrow\; & \frac{1}{2\ell}\left\langle Q_\rho(0)\,  (\mathcal W_{\mathfrak A})_{\phantom{\rho}\rho'}^{\rho}[0\leftarrow w]\,D_w^{\mathfrak A} \widetilde Q^{\rho'}(w) - \widetilde Q^{\rho'}(0)\,  (\mathcal W_{\mathfrak A})^{\phantom{\rho'}\rho}_{\rho'}[0\leftarrow w] \,D_w^{\mathfrak A} Q_\rho(w)\right\rangle \\
=\;&\frac{1}{2\pi\ell^2}\sum_{\rho}\; \partial_w \left( \frac{\theta'_1(0|\tau)}{\theta_4(\rho(\mathfrak A)|\tau)} \frac{\theta_4(-\frac{w}{2\pi} - \rho(\mathfrak A) | \tau)}{\theta_1(-\frac{w}{2\pi}|\tau)} \right)\;.
\end{align}
In the first line we omitted the antiholomorphic coordinates to avoid clutter. Upon removing the singular term proportional to $w^{-2}$, one can perform the limit $w\rightarrow 0$. One finds
\begin{equation}
\langle T_{\mathfrak A}(0)\rangle  = \frac{1}{24\pi^2\ell^2}\sum_\rho \left( - 3\frac{ \theta''_4(\rho(\mathfrak A)|\tau)}{\theta_4(\rho(\mathfrak A)|\tau)} +\frac{\theta'''_1(0|\tau)}{\theta'_1(0|\tau)}\right) \;.
\end{equation}
Using that the theta functions satisfy 
\begin{equation}
\label{theta-function-identity-1}
4\pi i \frac{\partial \theta_{j}}{\partial\tau}(z|\tau) = \frac{\partial^2 \theta_{j}}{\partial z^2}(z|\tau)\;, \qquad \text{for}\qquad j=1,2,3,4\;,
\end{equation}
we can trade the various double derivatives with respect to the first argument for derivatives with respect to $\tau$
\begin{equation}
\langle T_{\mathfrak A}(0)\rangle  = \frac{1}{2\pi \ell^2}\sum_\rho \left(- i\partial_\tau \ln \theta_4(\rho(\mathfrak A)|\tau) + \frac{i}{3}\partial_\tau \ln \theta'_1(0|\tau)\right) \;.
\end{equation}

Putting everything together, using that $\theta_1'(0,\tau) = 2 \pi\, \eta(\tau)^3$, we finally find that
\begin{equation}
\frac{\partial}{\partial \tau} \ln Z^{\text{HM}}(\mathfrak A,\tau) = 
\frac{\partial}{\partial \tau} \ln \left[\prod_\rho\left( (2\pi)^{\frac{1}{3}}\theta_4(\rho(\mathfrak A)|\tau)^{-1} \eta(\tau)\right)\right]\;.
\end{equation}
This differential equation is easily solved:
\begin{equation}\label{ZHMsoln}
Z^{\text{HM}}(\mathfrak A,\tau) = F(\mathfrak A)\;\prod_\rho \frac{\eta(\tau)}{\theta_4(\rho(\mathfrak A)|\tau)}\;,
\end{equation}
where $F$ is an undetermined function of $\mathfrak A$ (in which we also absorbed the factors of $2\pi$). All that is left to do, is prove that $F$ is in fact simply one. To show that it is independent of $\mathfrak A$, we can study the variation of $Z^{\text{HM}}(\mathfrak A,\tau)$ with respect to $\mathfrak A$. 

To avoid clutter, let us consider a single hypermultiplet with flavor symmetry fugacity $a$. Then one finds
\begin{equation}
\frac{\partial}{\partial a} Z^{\text{1HM}}(a,\tau) =  \frac{1}{i\pi\ell(\tau - \bar \tau)} \int[DQ D\widetilde Q] \left(\int_{T^2_{\tau}}  \sqrt{g}\ d^2z \ J(z)\right) \ e^{-\frac{1}{\pi\ell} \int_{T^2_{\tau}} \sqrt{g}\ d^2z\ \widetilde Q \left(\partial_{\bar z} - i \frac{a}{\bar\tau - \tau}\right)Q}\;,
\end{equation}
where $J(z)$ is the holomorphic current
\begin{equation}
J(z)\colonequals (\widetilde QQ)(z)\;.
\end{equation}
Following steps similar to the ones above, one finds
\begin{equation}\label{flavor-pde}
\frac{\partial}{\partial a} \ln Z^{\text{1HM}}(a,\tau) = - \frac{\theta_4'(a|\tau)}{\theta_4(a|\tau)}
\end{equation}
It is clear that for this equation to be satisfied by \eqref{ZHMsoln}, the function $F$ should be constant. We have no independent way to fix this constant, and simply choose it so that $Z^{\text{HM}}(\mathfrak A,\tau)$ starts off with unit coefficient. This completes the proof of \eqref{HMcheck}.


\section{Exact correlation functions on \texorpdfstring{$S^3\times S^1$}{S³xS¹} }\label{section:insertions}
The trace formula for the Schur index \eqref{SchurIndexDefn} can be further decorated with insertions of local operators. A local operator $\mathcal O$ acts as an endomorphism on the Hilbert space of states on $S^3$ and the trace will receive contributions from states $|\psi\rangle$ with nonzero overlap $\langle\psi|(-1)^F q^{E-R}\mathcal O|\psi\rangle$. To ensure the standard cancellations between bosonic and fermionic states that guarantee that long multiplets do not contribute to the index, the local operators should be preserved by the supercharge with respect to which the index is computed. In this paper, we take that supercharge to be a linear combination of the supercharges $\qq_1$ and $\qq_2$. The preserved operators are then precisely the operators constituting the associated chiral algebra.\footnote{This is not immediately obvious, because in general $\ker(\qq_1+\qq_2) \neq \ker(\qq_i)$. Note that $\qq_1+\qq_2$ is not nilpotent, but rather $(\qq_1+\qq_2)^2 = -\frac{1}{2}(r+2\mathcal M^\perp)$. Let us thus, without loss of generality, restrict to operators that lie in the kernel of $r+2\mathcal M^\perp$. On this subspace of operators, $\qq_1+\qq_2$ is nilpotent. Because exact insertions do not contribute in the trace, we can consider cohomologies instead of kernels, and a spectral sequence argument for the obvious double complex immediately shows that the cohomology of $\qq_1+\qq_2$ (on the restricted set of operators) equals the cohomology of either $\qq_i$.} 

Let us briefly remind the reader of the elementary chiral algebra fields on flat space in a Lagrangian theory \cite{Beem:2013sza}. For a generic gauge theory as above, involving a gauge group $G$ and a collection of hypermultiplets transforming in a representation $\mathfrak R$ of the combined gauge and flavor group, the chiral algebra fields are all words strung from\footnote{These elementary fields are, in the terminology of the introduction, ``transported'' versions of the hypermultiplet scalars $Q=q_{I=1,A=1}$ and $\widetilde Q = q_{I=1,A=2}$ and the gaugini $\lambda_{1+},\tilde \lambda_{1\dot +}$. On the torus we have already encountered the counterparts of transported $Q$ and $\widetilde Q$, see \eqref{defnQQtilde}, and shortly we will introduce the counterparts of the transported gaugini.}
\begin{itemize}
\item the symplectic boson pair $Q(z), \widetilde Q(z)$ transforming in $\mathfrak R,\mathfrak R^*$ respectively,
\item the small $(b,c)$-ghost pair $\partial_z c(z) = \lambda(z),\, b(z) = \tilde \lambda(z)$, transforming in the adjoint representation of the gauge group $G$,
\item the holomorphic derivative $\partial_z$,
\end{itemize}
subject to the condition that they lie in the cohomology of a certain nilpotent charge $Q_{\text{BRST}}$. While we refer the reader to the original paper \cite{Beem:2013sza} for the detailed definition of $Q_{\text{BRST}}$, we would like to recall in particular that it was shown there that $Q_{\text{BRST}}$ acts on gauge invariant objects made from the above ingredients as the on-shell variation under $\widetilde Q^2_{\dot -}$. Alternatively identifying the small $(b,c)$-ghost system as $b(z) = \lambda_z(z) , \partial_z c(z) = \tilde \lambda_z(z)$, one can similarly construct a BRST charge that acts on gauge invariant words as the on-shell supercharge $\widetilde Q_{1-}$. Recall that the on-shell action of a supercharge is obtained by imposing the auxiliary field equations, which for an $\mathcal N=2$ theory entails replacing the triplet of auxiliary fields $D_{IJ}$ with the moment map operator $\mu_{IJ}$. For Lagrangian theories, the latter is the adjoint-valued combination $(\mu_{IJ})_A^{\phantom{A}B}=q_{(IA}q_{J)}^B$.

In the path integral formulation of the index, we could consider insertions of any local operator.\footnote{More generally, the computation is also compatible with the insertion of non-local operators, in particular surface defects.} Generic insertions don't respect any of the supercharges in the supersymmetry algebra $\mathfrak{su}(1|1)\oplus \mathfrak {su}(1|1)$ preserved by the background. In particular, the independence of the result of exactly marginal couplings is no longer guaranteed and the interpretation as a decorated superconformal index is lost. Such generic insertions are obviously also incompatible with our localization computation. Our first task is thus to find out what, from the point of view of localization, the allowed insertions are. We will find that precisely the ($S^3\times S^1$ versions of the) chiral algebra fields discussed in the previous paragraph are localizable. Note that we will employ the vector multiplet action as supersymmetrically exact deformation action, and hence the requirement that the insertions be compatible with localization is tantamount to the statement that the result is independent of exactly marginal couplings.\footnote{This choice of localizing action is by no means necessary, but it simplifies matters as it removes the need to separately verify independence of exactly marginal couplings.} Once we have verified the set of localizable insertions, we will explain how to actually compute their correlation functions using localization.

\subsection{Localizable insertions}\label{subsec_localizableinsertions}
To find the most general insertions compatible with a localization computation, we revisit the key step in the localization argument \eqref{keyLocStep}, now in the presence of insertions. The question is about the $\mathfrak s$ independence of\footnote{To be more precise, when a dynamical vector multiplet is present one should deform the action with a $(\mathcal{Q} + \mathcal{Q}_B)$-exact term, where $\mathcal{Q}$ and $\mathcal{Q}_B$ denote a supercharge and the gauge-fixing BRST charge, respectively. We will of course only consider gauge invariant insertions $\mathcal O_{(\mathfrak s)}$, thus this slight imprecision is not important for the argument presented here.}
\begin{equation}
\langle \mathcal{O}_{(\mathfrak s)} \rangle_\mathfrak{s} = \frac{1}{Z} \int [D\Phi]\, \mathcal{O}_{(\mathfrak s)}\, e^{-S[\Phi] - \mathfrak{s} \mathcal{Q}\int V} \;,
\end{equation}
For complete generality, we allow the insertion $\mathcal O_{(\mathfrak s)}$ itself to have $\mathfrak s$ dependence. It will become clear shortly why it is important to allow for such dependence. The requirement that $\langle \mathcal{O}_{(\mathfrak s)} \rangle_\mathfrak{s}$ be independent of $\mathfrak s$ then reads
\begin{multline}\label{localizability}
\mathcal O_{(\mathfrak s)}\text{ is localizable} \quad \Longleftrightarrow \\
 0=\frac{d}{d\mathfrak{s}}\langle \mathcal{O}_{(\mathfrak s)}\rangle_\mathfrak{s} =  \frac{1}{Z} \int [D\Phi]\, \left( \frac{d}{d\mathfrak{s}}\mathcal{O}_{(\mathfrak{s})} +  (-1)^{F_{\mathcal O}}\, (\mathcal Q \mathcal O_{(\mathfrak s)})\, {\int V}\right)\, e^{-S[\Phi] - \mathfrak{s} \mathcal{Q}\int V} \;.
\end{multline}
Here $F_\mathcal{O}$ is the fermion number of the operator $\mathcal{O}_{(\mathfrak s)}$. It arises when we move the supercharge away from $\int V$. The standard statement inferred from this equation is 
\begin{equation}
\frac{d}{d\mathfrak{s}}\mathcal{O}_{(\mathfrak{s})}=0 \quad {\text{and}} \quad  \mathcal Q \mathcal O_{(\mathfrak s)}=0 \qquad \Longrightarrow \qquad \mathcal O \text{ is localizable}\;.
\end{equation}
It is quite obvious though, but has not been considered in the literature before, that even if an operator is not $\mathcal{Q}$-closed on its own, the localization principle remains perfectly valid as long as the path integral in the right-hand side of \eqref{localizability} vanishes. The task in any given localization computation is thus to look for all observables solving \eqref{localizability}.

Let us do so for the specific case of our localization computation. Concretely, we proceed in two steps. First, we localize the vector multiplet, while at the same time using the Yang-Mills action as $\mathcal Q$-exact deformation. Afterwards, we will localize the hypermultiplet action. Our initial deformed action thus reads
\begin{equation}\label{deformedaction}
S_{\text{deformed}} = S_\text{HM}+\mathfrak{s} S_\text{YM} = S_\text{HM}+\mathfrak{s} \mathcal{Q} \int \Xi
\end{equation}
where $\Xi = \frac{1}{2} (\Xi^- + \Xi^+) $ as defined in \eqref{Q-exactness-of-YM}. What insertions solve \eqref{localizability}? Recalling the definition of $Q$ and $\widetilde Q$ from \eqref{defnQQtilde}
\begin{equation}\label{defnQQtildebis}
Q \colonequals 2\pi e^{-\frac{\pi i}{4}}\sqrt{\ell}\ \big( e^{\frac{i\varphi}{2}} q_{11} - e^{-\frac{ i\varphi}{2}} q_{21}\big)\;, \qquad  
  \widetilde Q \colonequals 2\pi e^{-\frac{\pi i}{4}}\sqrt{\ell}\ \big(e^{\frac{i\varphi}{2}} q_{12} -  e^{- \frac{i\varphi}{2}} q_{22}\big) \;.
\end{equation}
and additionally introducing\footnote{Here $\sigma_z = - \frac{i\ell}{2}(\sigma_4 + i \sigma_1)$, and  $\tilde\sigma_z = - \frac{i\ell}{2}(\tilde\sigma_4 + i \tilde\sigma_1)$.}
\begin{equation}\label{defnlambdalambdatilde}
  \tilde \lambda_z \colonequals \frac{c_{\lambda}}{\sqrt{\ell}} (\xi^I \sigma_z \tilde \lambda_I)\;, \qquad \lambda_z \colonequals \frac{c_{\lambda}}{\sqrt{\ell}}(\tilde \xi^I \tilde\sigma_z \lambda_I) \;, 
\end{equation}
where $c_{\lambda} = 2\pi i \sqrt{2}$, we will show in the rest of this subsection that
\begin{claim}[\textit{Localizability}\label{localizablestuff}] \textit{Gauge invariant composites $\mathcal O_{(\mathfrak s)}$ built from the letters $Q$, $\widetilde Q$, $\sqrt{\mathfrak s}\lambda_z$, $\sqrt{\mathfrak s}\tilde \lambda_z$ and the covariant derivative $D_z$, and inserted on the torus $T^2_{\vartheta=0}$, such that $\left(\mathcal Q^{(\text{os})} \mathcal O_{(\mathfrak s)}\right)\big|_{T^2_{\vartheta=0}} = 0$ are localizable.}
\end{claim}

\noindent Here $\mathcal Q^{(\text{os})}$ denotes the on-shell supercharge $\mathcal Q$.\footnote{Recall from above that the on-shell action of a supercharge is obtained by imposing the auxiliary field equations, \ie{}, $D_{IJ}\rightarrow \mu_{IJ}$, in the transformation rules.}  In other words, in light of the discussion in the beginning of this section, we will show that precisely the torus counterparts of the chiral algebra fields surviving the BRST procedure associated to gauging, are localizable. For composites $\mathcal O$ built solely with bosonic letters, the above statement reduces to the standard assertion that the insertion should be $\mathcal Q$-closed. It is easy to verify that the only $\mathcal Q$-closed bosonic operators are built from $Q$ and $\widetilde Q$, as these letters vary to zero on the torus. Claim \ref{localizablestuff}, however, proposes a vast generalization of localizable insertions containing also fermionic letters and covariant derivatives. Finally, also note that we accompanied the fermionic letters with an explicit factor of $\sqrt{\mathfrak s}$.

To prove claim \ref{localizablestuff}, we have to verify that the composites satisfy the requirement in \eqref{localizability}. Let us make a few prelimary comments and observations. First, note that the operators $\mathcal O_{(\mathfrak s)}$ are not $\mathcal Q$-closed. Indeed, on the separate letters one finds (for insertion points on the torus $T^2_{\vartheta=0}$)
\begin{equation}\label{variationsOfLetters}
\begin{aligned}
&\mathcal Q Q =0\;, \quad && \mathcal Q (\sqrt{\mathfrak s}\lambda_z) = c_\lambda\sqrt{\frac{\mathfrak s}{\ell}}\,\mathfrak D_z \;,\quad  &&\mathcal Q D_z = -\frac{\sqrt{\ell}}{c_\lambda}(\lambda_z - \tilde \lambda_z)\;,\\
&\mathcal Q \widetilde Q =0\;,\quad &&\mathcal Q (\sqrt{\mathfrak s}\tilde \lambda_z) = c_\lambda\sqrt{\frac{\mathfrak s}{\ell}}\,\mathfrak D_z\;,&&
\end{aligned}
\end{equation}
where $\mathfrak D_z$ was defined in \eqref{def:Dz}.\footnote{Note that performing the $\mathcal Q$-variation of some composite does not necessarily commute with the restriction to the torus, in particular not when the operator contains derivatives. The reason is that the Killing spinors themselves carry explicit position dependence. Therefore, one should in general first perform the variation of the composite $\mathcal O_{(\mathfrak s)}$ as a four-dimensional object, and only then consider the restriction to the torus. In formulae, $\mathcal{Q}X|_{\vartheta = 0} \equiv (\mathcal{Q} X)|_{\vartheta = 0}$, but $\mathcal{Q} (D_\mu X|_{\vartheta = 0})$ is (bad) notation for $(D_\mu (\mathcal{Q} X))|_{\vartheta = 0}$. However, for the derivative in the $z$-direction, it is easy to verify that $(D_z \xi^I)|_{T^2} = (D_z\tilde \xi^I)_{T^2} = 0$, and thus the problem does not pose itself. \label{restrvar}} Second, as the appearance of on-shell supersymmetry variations in our claim already hints at, a crucial role will be played by the path integral over the auxiliary fields $D_{IJ}$, or equivalently, over the fields $\mathfrak D_z,\mathfrak D_{\bar z},\mathfrak D_3$, as defined in \eqref{def:Dz}. The terms containing auxiliary fields in the deformed action \eqref{deformedaction} -- which we denote as $S_{\text{deformed}}|_{_{D}}$ -- can be written more usefully as in \eqref{Dcontents}. The Gaussian path subintegral over these fields can be performed exactly, resulting in the one- and two-point functions
\begin{align}\label{auxFieldCorr1}
&\int [D\mathfrak D_z D\mathfrak D_{\bar z} D\mathfrak D_3]\ \mathfrak D_{a}(x)\; e^{-S_{\text{deformed}}|_{_{D}}} = \frac{i}{2\mathfrak s} \mu_a(x) \ Z|_{_D}(\mathfrak s,\mu)\;, \qquad \text{for }a=z,\bar z,3\\\label{auxFieldCorr2}
&\int [D\mathfrak D_z D\mathfrak D_{\bar z} D\mathfrak D_3]\ \mathfrak D_{z}(x) \mathfrak D_{\bar z}(y)\; e^{-S_{\text{deformed}}|_{_{D}}} = \big(-\frac{1}{4\mathfrak s^2} \mu_z(x) \mu_{\bar z}(y) -\frac{1}{\mathfrak s} \delta(x-y)\big) \ Z|_{_D}(\mathfrak s,\mu)\;,
\end{align}
where 
\begin{equation}
Z|_{_D}(\mathfrak s,\mu) = \int [D\mathfrak D_z D\mathfrak D_{\bar z} D\mathfrak D_3]\ e^{-S_{\text{deformed}}|_{_{D}}} \sim e^{\frac{1}{4\mathfrak s^2} \int d^{4}x \mu_a \mu_a}\;,
\end{equation}
up to the usual infinite ($\mathfrak s$-dependent) normalization factor. Third, the fermionic functional $\Xi$ is at most linear in the auxiliary fields $\mathfrak{D}_{a=z, \bar z, 3}$,
\begin{equation}\label{gaugeFixingFunctional}
\Xi = \Xi_0 + \Xi_z \mathfrak{D}_{\bar z} + \Xi_{\bar z} \mathfrak{D}_z + 2\Xi_3 \mathfrak{D}_3\;,
\end{equation}
where in particular (see appendix \ref{app:Q-exactness} for more details)
\begin{equation}\label{XiZexplicit}
\Xi_z =  - \frac{1}{2} \cos^2 \frac{\vartheta}{2}\Big[ (\tilde \xi^I \tilde \sigma_z \lambda_I) + ( \xi^I \sigma_z \tilde\lambda_I) \Big]\;.
\end{equation}
Note that on the torus $T^2_{\vartheta=0}$, it simplifies to $\Xi_z|_{T^2_{\vartheta=0}} = - \frac{\sqrt{\ell}}{2c_\lambda} (\lambda_z + \tilde \lambda_z)$. Finally, the actions $S_{\text{VM}}$ and $S_{\text{HM}}$ are invariant under the standard $U(1)_r$ symmetry. Correlation functions should thus obey the $U(1)_r$ selection rules.\footnote{It is obvious, but important to re-emphasize, that the symmetry acts only on the quantum fields and not on the Killing spinors.}

With these observations in mind, we set out to compute the path integral in the right-hand side of \eqref{localizability}. Let $\mathcal O_{(\mathfrak{s})}$ be a generic word built from the letters $Q,\widetilde Q,\sqrt{\mathfrak s}\lambda_z,\sqrt{\mathfrak s}\tilde \lambda_z$ and the covariant derivative $D_z$, and inserted on the torus $T^2_{\vartheta=0}$. Note that $U(1)_r$ selection rules impose that there be an equal number of $\sqrt{\mathfrak{s}}\lambda_z$ and $\sqrt{\mathfrak{s}}\tilde \lambda_z$ in $\mathcal O_{(\mathfrak{s})}$, or the correlation function $\langle\mathcal O_{(\mathfrak{s})}\rangle_{\mathfrak s}$ trivially vanishes. Let us denote the (even) fermion number of $\mathcal O_{(\mathfrak{s})}$ as $F_{\mathcal O}$. One then easily finds for the first term in the right-hand side of \eqref{localizability}
\begin{equation}\label{contributon1localizability}
\big\langle \frac{d}{d\mathfrak s}\mathcal{O}_{(\mathfrak s)}\big\rangle_\mathfrak{s} = \frac{F_{\mathcal O}}{2\mathfrak{s}} \left\langle\mathcal{O}_{(\mathfrak s)}\right\rangle_\mathfrak{s} \;.
\end{equation} 
Analyzing the second term requires some more work. We first act with the supercharge $\mathcal Q$ on $\mathcal{O}_{(\mathfrak s)}$ according to the graded Leibniz rule; its action on the individual letters was given in \eqref{variationsOfLetters}. Our next step is to perform the path subintegral over the auxiliary fields. There are two sources of auxiliary fields in the path integrand: the $\mathcal Q$-variation of the letters $\sqrt{\mathfrak{s}}\lambda_z$ and $\sqrt{\mathfrak{s}}\tilde \lambda_z$, and the fermionic functional $\Xi$. We thus encounter either an auxiliary one- or two-point function. From \eqref{auxFieldCorr1}, it is clear that the one-point functions can be implemented simply by the replacement $\mathfrak D_a \rightarrow \mu_a$ in both the integrand and the action. The two-point functions result in a term that can be obtained by applying the same replacement rule, and a term from the delta-function in \eqref{auxFieldCorr2}. Let us look at this second term in some more detail. The two-point function arises when $\mathcal Q$ hits a fermionic letter in $\mathcal O_{(\mathfrak s)}$, producing a $\mathfrak D_z$ in combination with the $\mathfrak{D}_{\bar z}$ term in $\Xi$. A term in the path integrand exhibiting this occurrence looks like
\begin{equation}
(-1)^{F_\mathcal{O}+L_\lambda - 1}[ \ldots \mathcal{Q}(\sqrt{\mathfrak{s}}\lambda_z(z_i))\ldots ](\Xi_z\mathfrak{D}_{\bar z})(x) =
  (-1)^{F_\mathcal{O}+L_\lambda -1}\big[ \ldots c_\lambda\sqrt{\frac{\mathfrak{s}}{\ell}}\mathfrak{D}_z(z_i) \ldots \big](\Xi_z\mathfrak{D}_{\bar z})(x). \nn 
\end{equation}
Here we used that $\mathcal{Q}$ is a fermionic charge and it produces factors of minus one as it hops over fermionic letters. The integer $L_\lambda$ denotes the position of the particular letter $\lambda_z(z_i)$ in $\mathcal{O}_{(\mathfrak s)}$ among the fermionic letters and counted from the left. We have left implicit the integration over $x$. The delta-function term of the auxiliary two-point function can be easily integrated over space, resulting in
\begin{equation}
  (-1)^{F_{\mathcal O } + L_\lambda - 1}\big(  - \frac{1}{\mathfrak{s}}  \big)\Big[\ldots (1) \ldots\Big]\left(-\frac{1}{2}\right)(\sqrt{\mathfrak{s}}\lambda_z(z_i) + \sqrt{\mathfrak{s}}\tilde \lambda_z(z_i)) \ Z|_{_D}(\mathfrak s,\mu) \; ,
\end{equation}
where we used the observation made below \eqref{XiZexplicit} to simplify $\Xi_z$. We kept the order of all letters and left an explicit $(1)$ at the original location of $\mathcal Q\lambda_z(z_i)$. The $U(1)_r$ selection rule allows us to remove the $\tilde\lambda_z(z_i)$ in the last parenthesis, and finally we move the $\lambda_z(z_i)$ sitting at the end back to the location of the symbol $(1)$. Doing so, we have resurrected the $\lambda_z(z_i)$ in its original position, while also creating a factor $(- \frac{1}{2})(-\frac{1}{\mathfrak{s}})(-1)^{F_{\mathcal O}+L_\lambda - 1 + (F_{\mathcal O}-L_\lambda)} Z|_{_D}(\mathfrak s,\mu) = - \frac{1}{2\mathfrak{s}} Z|_{_D}(\mathfrak s,\mu)$. A similar process takes place for insertions of $\sqrt{\mathfrak{s}}\tilde\lambda_z(z_i)$. If the fermionic letter is acted on by covariant derivatives, one can proceed as follows to analyze the delta-function term of the auxiliary two-point function. We will focus on the case of a single derivative, but the analysis can be easily extended. The expression
\begin{equation}
  (-1)^{F_{\mathcal O}+L_\lambda - 1}\int d^4 x\sqrt{g}\, \big[\ldots c_\lambda \sqrt{\frac{\mathfrak{s}}{\ell}}D_{z_i}\mathfrak{D}_z(z_i) \ldots\big]\Xi_z(x)\mathfrak{D}_{\bar z}(x) \; ,
\end{equation}
becomes upon using the delta-function term in the auxiliary field two-point function,
\begin{align}
  & (-1)^{F_{\mathcal O}+L_\lambda - 1}\left(- \frac{1}{2}\right)\left(-\frac{1}{\mathfrak{s}}\right)\ Z|_{_D}(\mathfrak s,\mu)\Big[\ldots(1)\ldots\Big]\int d^4 x \sqrt{g}\, D_{z_i} \delta(z_i - x) \cos^2\frac{\theta}{2}\sqrt{\mathfrak{s}}\lambda_z(x) \nn \\
  = & (-1)^{F_{\mathcal O}+L_\lambda - 1}\left(- \frac{1}{2}\right)\left(-\frac{1}{\mathfrak{s}}\right)\ Z|_{_D}(\mathfrak s,\mu)\Big[\ldots(1)\ldots\Big]\int d^4 x \sqrt{g}\, \delta(z_i - x) \cos^2\frac{\theta}{2}\sqrt{\mathfrak{s}}D_{z_x}\lambda_z(x) \nn\\
  =& - \frac{1}{2\mathfrak{s}} \Big[\ldots(\sqrt{\mathfrak{s}}D_{z_i} \lambda_z(z_i))\ldots\Big]\, Z|_{_D}(\mathfrak s,\mu)\ ,
\end{align}
restoring $D_{z_i}\lambda_z(z_i)$ in its original position in $\mathcal{O}(z_i)$, together with the same factor as before. Here $z_x$ denotes $\varphi_x + \tau t_x$.\footnote{Some more details are as follows. First we used that $D_{z_i} \delta(z_i - x) = - D_{z_x}\delta(z_i - x)$, and subsequently we performed the partial integration which leads to a second minus sign.} In total, this type of restoration occurs for each fermionic letter, which is $F_{\mathcal O}$ times in total. Since $\mathcal O_{(\mathfrak s)}$ doesn't contain auxiliary fields itself, we are thus left with $\big\langle -\frac{F}{2\mathfrak{s}} \mathcal{O}(z_i)\big\rangle_\mathfrak{s}$ from all delta-function terms of the auxiliary two-point function. This contribution precisely cancels against \eqref{contributon1localizability}.

At this point, we are left with the conclusion that
\begin{equation}
\frac{d}{d\mathfrak{s}}\langle \mathcal{O}_{(\mathfrak s)} \rangle_\mathfrak{s}  \propto \int \sqrt{g}d^4 x\Big\langle \big[\mathcal{Q}^\text{(os)}\mathcal{O}_{(\mathfrak s)} \big] \Xi(x) \Big \rangle_\mathfrak{s}\;,
\end{equation}
where as before $\mathcal{Q}^\text{(os)}$ denotes the on-shell action of the supercharge $\mathcal Q$. This result proves our claim.\footnote{Implicitly we have already used that the only bosonic letters that will appear in $\mathcal{Q}^\text{(os)}$-closed objects are $Q$ and $\widetilde Q$.} Note that we can only truly declare success after also localizing the hypermultiplet. We will postpone that localization computation to subsection \ref{HMrevisited}, and first analyze the result of the vector multiplet localization in the presence of insertions. However, the hypermultiplet localization will turn out to be trivially compatible with the insertions satisfying $\mathcal{Q}^\text{(os)}\mathcal{O}_{(\mathfrak s)} =0$.

\subsection{Localizing fermionic insertions}
In the previous subsection we have shown that certain fermionic insertions are compatible with localization. At first sight, the notion of localizing fermionic observables seems puzzling, since naively the localization computation instructs one to evaluate the insertions on the  purely bosonic localization locus.\footnote{Some computations feature fermionic zero modes which should be considered part of the localization locus. The integrations over said zero-modes need to be saturated appropriately, see for example \cite{Witten:1988ze,Benini:2013xpa}. We focus on localization computations that do not feature fermionic zero-modes.} Let us thus explain how to properly incorporate fermionic insertions in a localization computation.

As we reviewed above, to perform a localization computation, one deforms the original physical action by a deformation parameter $\mathfrak s$ times a $\mathcal{Q}$-exact (and simultaneously $\mathcal{Q}$-closed) localizing action $S_\text{def}$, with the property that its bosonic piece $S_\text{def}^\text{B}$ is positive semi-definite. Let us consider the insertion of an operator $\mathcal{O}_{(\mathfrak s)}$ which we split in a purely bosonic part and a part made out of fermionic fundamental fields: $\mathcal{O}_{(\mathfrak s)} = \mathcal{O}^\text{B}\mathcal{O}^\text{F}_{(\mathfrak s)}$. If $\langle \mathcal{O}_{(\mathfrak s)}\rangle_\mathfrak{s}$ can be shown to be independent of $\mathfrak s$, one has the freedom to send $\mathfrak s$ to infinity $\mathfrak{s} \to +\infty$. The bosonic part of the deformation action, which enters the path integral as $e^{-\mathfrak s S_\text{def}^\text{B}}$ forces the bosonic path integral to localize onto the localization locus defined by the zeros of the deformation Lagrangian. Let us parameterize this locus by, say, $\mathfrak b$. At this point one can perform the path integration over the bosonic fields. Following the standard steps, one simply has to evaluate the bosonic insertion $\mathcal{O}^\text{B}$ and the original classical action $S_{\text{cl}}$ on the localization locus, while the quadratic expansion of $\mathfrak{s} S_\text{def}^\text{B}$ around the locus  provides the bosonic one-loop determinant $\big(\det (\mathfrak{s}\, \mathcal D_\text{B}(\mathfrak{b}))\big)^{-1}$. Here $\mathcal D_\text{B}(\mathfrak{b})$ is the operator describing the quadratic fluctuations. At this point, the result of the path integral reads
\begin{equation}
\langle \mathcal{O}_{(\mathfrak s)}\rangle_\mathfrak{s} = \frac{1}{Z}\int [d\mathfrak{b}] \Big[ \mathcal{O}^\text{B}|_{\mathfrak b}\  e^{- S_\text{cl}|_\mathfrak{b}}\ \frac{1}{\det (\mathfrak{s}\, \mathcal{D}_\text{B}(\mathfrak{b}))} \int [D\Lambda]\ \mathcal{O}^\text{F}(\sqrt \mathfrak{s}\Lambda)|_{\mathfrak b}\ e^{ - \mathfrak{s} S_\text{def}^\text{F}[\Lambda]|_\mathfrak {b}} \Big]\;,
\end{equation}
where ``$|_{\mathfrak b}$'' denotes the evaluation on the localization locus, and $\Lambda$ collectively denotes the fermionic fields. Note that the fermionic deformation action, evaluated on the bosonic localization locus, is Gaussian.\footnote{Also note that the fermionic piece of the original action is suppressed with respect to the deformation action.} This Gaussian integral can be performed straightforwardly, even in the presence of the fermionic insertions $\mathcal{O}^\text{F}$: one can simply evaluate it using the relevant propagators and knowledge of the path integral in the absence of insertions. The latter will of course evaluate to the fermionic one-loop determinant $\det (\mathfrak s\mathcal{D}_\text{F}(\mathfrak{b}))$. The ratio of fermionic and bosonic one-loop determinants reproduces the standard total one-loop determinant (and is independent of $\mathfrak s$). Note that the quadratic action governing the fermionic fields still has an explicit $\mathfrak s$ up front and that the fermionic insertions have an explicit $\sqrt{\mathfrak s}$ to ensure localizability.\footnote{We have shown this explicit dependence only in the concrete case of our localization computation, but believe it to be a general feature.} Making a change of variables from $\Lambda$ to $\mathfrak s^{-1/2} \Lambda$, and explicitly using the localizability of the correlator of interest, in other words, its $\mathfrak s$ independence, we can simplify the above expression slightly by using that the Jacobian must cancel against the $\mathfrak s$ dependence of the bosonic one-loop determinant:
\begin{equation}\label{VMfermionicLoc}
\langle \mathcal{O}_{(\mathfrak s)}\rangle_\mathfrak{s} = \frac{1}{Z}\int [d\mathfrak{b}] \Big[ \mathcal{O}^\text{B}|_{\mathfrak b}\  e^{- S_\text{cl}|_\mathfrak{b}}\ \frac{1}{\det \mathcal{D}_\text{B}(\mathfrak{b})} \int [D\Lambda]\ \mathcal{O}^\text{F}(\Lambda)|_{\mathfrak b}\ e^{ - S_\text{def}^\text{F}[\Lambda]|_\mathfrak {b}} \Big]\;.
\end{equation}

\subsection{Schur correlation functions on \texorpdfstring{$S^3\times S^1$}{S³xS¹} }
Let us return to our localization computation on $S^3\times S^1$. At the end of section \ref{subsec_localizableinsertions}, we achieved a detailed understanding of the localizable insertions with respect to the vector multiplet localization. In the previous subsection, we subsequently explained how fermionic insertions can be incorporated in a general localization computation. Let us now combine these two pieces of information to complete the localization with operator insertions on $S^3\times S^1$. Let us start by noting that we have not yet explicitly localized the hypermultiplet. For computational purposes, it is however sufficient to observe that on any given fixed (off-shell) vector multiplet configuration, the hypermultiplet action is quadratic. Thus, after the vector multiplet localization, the entire path integral has collapsed to a Gaussian path integral, coupled to an integral over the vector multiplet localization locus. The Gaussian path integral comprises the integral over fermions of the previous subsection, and the path integral over hypermultiplet fields. In the presence of various insertions, the quadratic path integral is easily evaluated, most conveniently by evaluating the relevant propagators and using Wick's theorem. Let us start by evaluating these propagators.

\subsubsection{Fermionic propagator \texorpdfstring{$\langle  \tilde \lambda_z(z) \lambda_z(w)\rangle$}{}} 
As explained in the previous subsection, the propagator for the fermionic insertions is computed from the fermionic part of the deformation action. In our case, that is simply the Yang-Mills action itself, and the relevant term is $2i \tilde\lambda^I\tilde\sigma^\mu D_\mu \lambda_I$. Recall that the covariant derivative contains the $U(1)_r$ and $SU(2)_R$ connections, given in \eqref{SU2RandU1rconnection}, and also the gauge connection evaluated on the localization locus $A = \mathfrak b dt$. 

Let us start by computing the propagator $(G_{IJ})_\alpha{^{\dot \beta}}(x) \colonequals \langle\tilde \lambda_J^{\dot \beta}(0)\lambda_{I\alpha}(x)\rangle$, where we have suppressed the adjoint gauge indices. Since we don't expect the final answer for $\langle \tilde \lambda_z(0)  \lambda_z(z)\rangle$ to depend on $\beta_1$ or $\beta_2$, we can choose to set them to zero from the start: $\beta_1=\beta_2=0$. For simplicity, we will also choose $\tau=i$, \ie{}, we consider the round three-sphere. It will be clear how to reinstate generic $\tau$ later on. On this geometry, the Green's function  $(G_{IJ})_\alpha{^{\dot \beta}}$ can be obtained straightforwardly in two steps. First, we perform a Weyl transformation to $S^3\times \mathbb R$ of the flat space Green's function of the standard Dirac operator.\footnote{One also needs to perform a change of coordinates and perform a frame rotation to align the Weyl transformed vielbein with our choice of vielbein in \eqref{vielbein}.} Upon additionally incorporating the presence of the $SU(2)_R$ connection and gauge connection, we find the nonzero components
\begin{align}
(G_{12}^{S^3\times \mathbb R})_\alpha^{\phantom{\alpha}\dot \beta}(x)
  = & \ -\frac{i}{ (2\pi)^2 \ell^3}\diag(e^{i \varpi(\mathfrak b)t}) \frac{e^{ - \frac{t}{2}}e^{\frac{1}{2}t-\frac{i}{2}(\theta + \varphi + \chi)}}{2(e^{-t}+e^{t}- 2 \cos\theta \cos \varphi)^2}\\
  & \ \times 
  \left(
  \begin{array}{cc}
   e^t-e^{i (\theta +\varphi )}+e^{i (\theta +\chi )}-e^{t+i (\varphi +\chi )} & -e^t+e^{i (\theta +\varphi )}+e^{i (\theta +\chi )}-e^{t+i (\varphi +\chi )} \\
   e^{t+i \theta }-e^{i \varphi }-e^{i \chi }+e^{t+i (\theta +\varphi +\chi )} & -e^{t+i \theta }+e^{i \varphi }-e^{i \chi }+e^{t+i (\theta +\varphi +\chi )} \\
  \end{array}
  \right),\nn
\end{align}
and $(G_{21}^{S^3\times \mathbb R})_\alpha{^{\dot \beta}}(x) = - e^{+ t} (G_{12})_\alpha{^{\dot \beta}}(x)$. The propagator is diagonal in its gauge indices,\footnote{More precisely, we mean that it is proportional to the Killing form on the Lie algebra.} with the diagonal elements given in terms of the action of the corresponding generator $\varpi \in \mathfrak g$ on the Cartan element $\mathfrak b$. The second step is to compactify $\mathbb R$ to $S^1$. We will only do so for the combination of our actual interest and with insertion points on the submanifold $S^1_{\vartheta=0}\times \mathbb R$ defined by $\vartheta=0$
\begin{align}
  \left(\frac{c_\lambda}{\sqrt{\ell}}\right)^2\big[\tilde \xi^I_{\dot \alpha}(\tilde \sigma_z)^{\dot\alpha \alpha} \big](x)\big|_{\vartheta_x \to 0}\ & \Big[\xi^{J\beta} (\sigma_z)_{\beta \dot \beta} \Big](0) \ (G_{IJ}^{S^3\times \mathbb R})_\alpha{^{\dot \beta}}(x)\big|_{\vartheta_x \to 0} 
  = \frac{i}{4 \ell^2}\frac{\diag(e^{i \varpi(\mathfrak b)t})}{\sin^2 \frac{\varphi + i t}{2}}\;, \nn
\end{align}
where we filled in $c_\lambda = 2\pi i \sqrt{2}$. Note that the Killing spinors $\xi, \tilde \xi$ on $S^3 \times S^1$ are independent of $t$, hence it makes sense to have them defined on the uncompactified $S^3 \times \mathbb{R}$. Also observe that the denominator of the answer only depends on the holomorphic combination $\varphi + i t$. It is clear that for generic $\tau$, we would have found instead $\varphi + \tau t = z$. The compactification of $\mathbb{R}$ to $S^1$ is implemented by enforcing periodicity $t \sim t + 2\pi n$. We thus consider, temporarily omitting prefactors to avoid clutter,
\begin{equation}
\sum_{n = -\infty}^{+\infty} \frac{\diag(e^{i \varpi(\mathfrak b)(t+2\pi n})}{\sin^2 ( \frac{\varphi + \tau t}{2} + n \pi \tau)} = \diag\left(e^{i \varpi(\mathfrak b)t} \sum_{n = -\infty}^{+\infty} \frac{e^{2 \pi i n\varpi(\mathfrak b)}}{\sin^2 ( \frac{\varphi + \tau t}{2} + n \pi \tau)}\right)\;.
\end{equation}
To perform this sum, let us start by observing that $\sum\limits_{m \in \mathbb{Z}} (\frac{a}{2} + m\pi)^{-2} = \sin^{-2} \frac{a}{2}$, and write
\begin{equation}
\diag\left(e^{i \varpi(\mathfrak b)t} \sum_{n = -\infty}^{+\infty} \frac{e^{2 \pi i n\varpi(\mathfrak b)}}{\sin^2 ( \frac{\varphi + \tau t}{2} + n \pi \tau)} \right)= \diag\left(e^{i \varpi(\mathfrak b)t} \sum_{m,n = -\infty}^{+\infty} \frac{e^{2 \pi i n\varpi(\mathfrak b)}}{( \frac{\varphi + \tau t}{2} + m \pi + n \pi \tau)^2}\right)\;.
\end{equation}
The latter sum can be recognized as the definition of the twisted Weierstrass function $P_2$, see \cite{tuite2007torus} for more details, or appendix \ref{specialfuctions} for a brief summary. Using properties proved in \cite{tuite2007torus}, it can be written in terms of derivatives of standard Jacobi theta functions as (see again appendix \ref{specialfuctions} for more details)
\begin{equation}
\sum_{m, n = -\infty}^{+\infty} \frac{e^{2\pi i n \varpi(\mathfrak{b})}}{(\frac{z}{2} + m \pi + n \pi \tau)^2} = 4 P_2 \left[\begin{smallmatrix}  e^{-2\pi i \varpi(\mathfrak{b})} \\ 1 \end{smallmatrix}\right](z|\tau) = 4i\, \eta(\tau)^3\partial_z
    \frac{\theta_{1}(\frac{ z }{2\pi} + \varpi(\mathfrak{b})|\tau)}{\theta_{1}(\varpi(\mathfrak{b})|\tau) \ \theta_{1}(\frac{ z }{ 2\pi  } |\tau)}\ ,
\end{equation}
In the end, the propagator reads
\begin{equation}\label{fermionicprop}
  \langle \tilde\lambda_z(0) \lambda_z(z,\bar z) \rangle =
 - \frac{1}{\ell^2} \eta(\tau)^3 \diag\left(e^{i \varpi(\mathfrak{b})\frac{z-\bar z}{\tau - \bar \tau}} \partial_z\bigg[   \frac{\theta_1(\varpi(\mathfrak{b}) + \frac{z}{2\pi} |\tau)}{\theta_1(\frac{z}{2\pi} |\tau)\theta_1(\varpi(\mathfrak{b})|\tau)}   \bigg]\right)\;.
\end{equation}
Despite appearances, this expression has a smooth limit for $\varpi \rightarrow 0$, which is of course relevant for the propagator of Cartan components. Concretely, it is given by
\begin{equation}
\lim_{\varpi\rightarrow 0}
  \frac{-1}{\ell^2}
  \eta(\tau)^3 e^{i \varpi(\mathfrak{b})\frac{z-\bar z}{\tau - \bar \tau}}
  \partial_z\bigg[
    \frac{\theta_1(\varpi(\mathfrak{b}) + \frac{z}{2\pi} |\tau)}{\theta_1(\frac{z}{2\pi} |\tau)\theta_1(\varpi(\mathfrak{b})|\tau)}
  \bigg]
=  -\frac{1}{\ell^2} \partial_z
  \bigg[\frac{\partial_z \theta_1(\frac{z}{2\pi}|\tau)}{ \theta_1(\frac{z}{2\pi}|\tau)} \bigg]  \; .
\end{equation}
Also note that \eqref{fermionicprop} is doubly periodic in $z$, as it should be.

\subsubsection{Bosonic propagator \texorpdfstring{$\langle  Q(z) \widetilde Q(w)\rangle$}{}} 
To compute $\langle  Q(z) \widetilde Q(w)\rangle$, we can follow similar steps as in the subsection above using the relevant terms in the hypermultiplet action \eqref{HM-Lagr}, $\frac{1}{2}{D_\mu }{q^{IA}}{D^\mu }{q_{IA}}+\frac{\mathcal R}{12}{q^{IA}}{q_{IA}}$, where the covariant derivatives contain the $SU(2)_R$ connection and the gauge connection evaluated on the dynamical and background vector multiplet localization locus $A=\mathfrak A dt$. Without presenting the details, it won't come as a surprise, and in fact will be a direct consequence of the additional hypermultiplet localization computation of subsection \ref{HMrevisited}, that the result precisely agrees with the propagator in \eqref{propagator}. For convenience, we reproduce it here:
\begin{equation}\label{bosonicprop}
\langle\widetilde Q^{\rho'}(0)\, Q_\rho(z,\bar z) \rangle=
 - \frac{i}{(2\pi)^3\ell^2}e^{- i\frac{\rho(\mathfrak{A})}{ \tau- \bar\tau} (\bar z - z)} \frac{\theta'_1(0|\tau)}{\theta_4(\rho(\mathfrak A)|\tau)} \frac{\theta_4(\frac{z}{2\pi} + \rho(\mathfrak A) | \tau)}{\theta_1(\frac{z}{2\pi}|\tau)}\; \delta_\rho^{\rho'}\; .
\end{equation}

\subsubsection{Result of localization with insertions}
Putting everything together, we find that a correlator of words strung from the letters $Q,\widetilde Q,\lambda_z,\tilde \lambda_z$ and the covariant derivative $D_z$, inserted on the torus $T^2_{\vartheta=0}$, is computable via localization as
\begin{equation}
\big\langle \prod_i \mathcal{O}_i(z_i) \big\rangle =  \frac{1}{|\mathcal W|} \oint \prod_{j = 1}^{\rank \mathfrak{g}} \frac{db_j}{2\pi i b_j}     Z^\text{HM}(\mathfrak{A}, \tau) Z_\text{1-loop}^\text{VM}(\mathfrak{b}, \tau)  \langle \prod_i \mathcal{O}_i(z_i) \rangle_{\text{GT}}\;,
\end{equation}
where we used the vector multiplet one-loop determinant as given in \eqref{oneloopVM} and the partition function of the symplectic boson given in \eqref{HMcheck}. In other words, in the absence of insertions, we simply recover \eqref{finalresultOLD}. The symbol $\langle \ldots \rangle_{\text{GT}}$ denotes the evaluation of the correlator in the relevant Gaussian theories, \ie{}, using the propagators \eqref{fermionicprop} and \eqref{bosonicprop}.

\subsection{Two-dimensional Gaussian theories}\label{HMrevisited}
Although we have already obtained a full computational understanding of correlators in the previous subsection, we can still choose to complete the localization computation of the hypermultiplet theory. After having reduced the vector multiplet path integral to an integral over the (bosonic) localization locus, possibly having evaluated correlation functions of insertions of gaugini, the remaining quantum field theory describing the dynamics of the hypermultiplet -- whose action is the original hypermultiplet action evaluated on the vector multiplet localization locus -- still has a supersymmetry. It is generated by $\mathcal Q|_{\mathfrak b}$, \ie{}, the transformation rules are those of $\mathcal Q$ evaluated on the vector multiplet localization locus. We can add the standard localization term $\mathfrak t \mathcal Q|_{\mathfrak b} V_{\text{HM}}$ to the hypermultiplet action, where $V_{\text{HM}}$ can be chosen as in \eqref{standarddefferm}. The remaining letters after the vector multiplet path integral is fully localized are just $Q,\widetilde Q$ and $D_z^{\mathfrak A}$. The operators built from these letters are trivially closed under $Q|_{\mathfrak b}$. The insertion of these $\mathcal{Q}|_{\mathfrak b}$-closed observables in the path integral propagates through the localization computation presented in section \ref{section:localizingSchur} in the obvious manner. Ultimately, the localization turns the hypermultiplet action into \eqref{HMactionSB}, directly describing the dynamics of the fields $Q$ and $\widetilde Q$. This theory is Gaussian, and its propagator was already presented in \eqref{propagator}, and recovered just now in \eqref{bosonicprop}.

A natural question to ask is whether we can identify, and preferentially derive from first principles, the two-dimensional theory governing the dynamics of $\lambda_z$ and $\tilde \lambda_z$. We expect the answer to be the small $(b,c)$ ghost system. A direct derivation thereof is unclear at the moment, but we will present a schematic analysis momentarily. Let us first remark that it of course passes all possible \textit{a posteriori} tests. For example, the quadratically divergent short-distance behavior and double-periodicity properties of the propagator \eqref{fermionicprop} correctly reproduce the expected behavior of the small $(b,c)$ ghost system. Furthermore, it is manifestly the case that the propagator is related to the propagator of the $(b,c)$ ghost system of weight $(1,0)$ by an additional derivative.\footnote{Note however that the propagator of the $(b,c)$ ghost system is slightly subtle due to the zero mode, so we do not claim a direct equality.} What's more, the graded torus partition function of the small $(b,c)$ ghost system trivially reproduces the one-loop determinant of \eqref{oneloopVM} (up to the four-dimensional gauge-fixing ghosts' contribution). Moreover, identifying the stress tensor of the system in the standard manner as $\lambda_z\tilde\lambda_z$, one can reconfirm this fact, completely similarly to the computation in subsection \ref{HMoneloopcomputation}, by solving the differential equation obtained from varying the two-dimensional theory with respect to $\tau$ and using the propagator to compute the stress tensor one-point function.

Let us now sketch the schematic picture of how to go about the analysis. Recall the structure of the result of the vector multiplet localization of equation \eqref{VMfermionicLoc}, and let us expand the fermionic fields in ``eigenmodes'' $\Lambda_n, \tilde\Lambda_n$ of the operator $\mathcal D_{\text{F}}$. It is intuitively clear that the fermionic letters contributing to localizable insertions, \ie{}, the combinations $\lambda_z$ and $\tilde\lambda_z$ inserted at a point on the torus $T^2$, are spanned by a subset of (linear combinations of) fermionic modes. All complementary modes can be integrated out. Our claim is that the product of eigenvalues of the integrated modes cancels the one-loop determinant of bosonic fluctuations. The underlying reason is that the spectrum of fluctuations around the localization locus organizes itself in representations of the $\mathfrak{su}(1|1)$ algebra generated by the localizing supercharge $\mathcal Q$, its Hermitian conjugate $\mathcal Q^\dagger$ and their anticommutator. The bosonic and fermionic modes whose contributions cancel, reside in long representations of this $\mathfrak{su}(1|1)$, and we expect all bosonic vector multiplet modes to indeed enter a long multiplet.\footnote{These cancellations are similar to the cancellations that take place in the superconformal index as a trace over the Hilbert space of states. It is also reminiscent of the ``missing mode'' analysis in \cite{Hama:2011ea}.} The theory of interest is thus described by the remaining fermionic modes.

\subsection{Application: Schur correlation functions in \texorpdfstring{$SU(2)\ \mathcal N=4$}{SU(2) N=4} SYM}
Let us apply the formalism developed in the previous several subsections to compute various Schur correlation functions in four-dimensional $\mathcal N=4$ super Yang-Mills theory with gauge group $SU(2)$. We will treat the theory as an $\mathcal N=2$ superconformal field theory. This theory possesses an $SU(2)_F$ flavor symmetry, which is the commutant of the $\mathcal N=2$ $R$-symmetry inside the full $SU(4)$ $R$-symmetry of the $\mathcal N=4$ theory. The elementary letters from which we should string gauge-invariant $\mathcal Q^{(\text{os})}$-closed words, are $Q^i, \lambda_z, \tilde \lambda_z$, all taking value in the adjoint representation of the gauge group $SU(2)$, and the covariant derivative $D_z$. Here the index $i$ denotes a fundemantal index of the flavor symmetry $SU(2)_F$.\footnote{In terms of the notation used above, $Q^1 = Q$ and $Q^2=\widetilde Q$.}  In \cite{Beem:2013sza}, it was conjectured that the full set of $Q^{(\text{os})}$-closed words is strongly generated by\footnote{All other words can be obtained by taking (normal ordered) products of (derivatives of) these words.}
\begin{align}
&T_{2d} \colonequals \tr\left(-\frac{1}{2\ell} Q^i D_z^{\mathfrak A} Q^j \epsilon_{ij} - \tilde\lambda_z \lambda_z \right)\;,\qquad && G^i \colonequals \sqrt{2} \tr Q^i \lambda_z\;,\\ 
&J^{(ij)} \colonequals -\frac{1}{2}\tr Q^i Q^j\;,\qquad  &&\tilde G^i \colonequals - \sqrt{2} \tr Q^i \tilde\lambda_z\;.
\end{align}
On flat space, the correlators of these objects were shown to define the small $\mathcal N=4$ superconformal algebra of central charge $c_{\text{2d}}=-9$. The two-dimensional interpretation of the various words is clear: one finds the stress tensor $T_{2d}$, a triplet of affine $\widehat{\mathfrak{su}(2)}_F$ currents $J^{(ij)}$ at level $k_{\text{2d}}=-\frac{3}{2}$, and a pair of supercurrents in the doublet of $\mathfrak{su}(2)$. Also recall that for this central charge, the stress tensor is cohomologous to the Sugawara stress tensor of the $SU(2)_F$ current
\begin{equation}
T_{\text{Sug.}} \colonequals \frac{1}{2 (k_{\text{2d}}+h^\vee)} J^{ij}J^{kl}\epsilon_{ik}\epsilon_{lj} = J^{ij}J^{kl}\epsilon_{ik}\epsilon_{lj}\;, \qquad \text{and} \qquad T_{2d} \equiv T_{\text{Sug.}}\;.
\end{equation}

Let us start by reminding the reader of the index of $\mathcal N=4$ super Yang-Mills with gauge group $SU(2)$. It is computed by the contour integral
\begin{align}
&I_{\mathcal N=4}^{(SU(2))}(q;a) = \frac{1}{2}\oint \frac{db}{2\pi i b}\, Z_{\text{1-loop}}^{\text{VM}}(\mathfrak b, \tau)\,Z^{\text{HM}}(\mathfrak a,\mathfrak b,\tau)\\ 
&=- \frac{1}{2}\oint \frac{db}{2\pi i b} \theta_1(2\mathfrak b|\tau)\theta_1(-2\mathfrak b|\tau) \frac{\eta(\tau)^3}{\theta_4(2\mathfrak b + \mathfrak a|\tau)\theta_4(-2\mathfrak b + \mathfrak a|\tau)\theta_4(\mathfrak a|\tau)} \\
&= q^{\frac{3}{8}} \Big(1+\chi_{\mathbf{3}}(a) q - 2 \chi_{\mathbf{2}}(a)q^\frac{3}{2} + (\chi_{\mathbf 1}(a) + \chi_{\mathbf{3}}(a) + \chi_\mathbf{5}(a))q^2 -2(\chi_{\mathbf{2}}(a)+ \chi_\mathbf{4}(a))q^{\frac{5}{2}} + \ldots\Big)\;.
\end{align}
where as before $a=e^{2\pi i \mathfrak a}$ and $b=e^{2\pi i \mathfrak b}$, and we wrote the $q$-expansion of the result in terms of characters of $SU(2)_F$ representations. 

Let us now compute explicitly various interesting one-point functions.\footnote{There is of course no obstruction to computing any arbitrary $n$-point correlator.} We start with the one-point function of the stress tensor $T_{2d}$. Our localization results show that it is computed by\footnote{We omit the position of the insertion, as the one-point function is independent thereof.}
\begin{equation}\label{T2d1pt}
\langle T_\text{2d}\rangle = \frac{1}{I_{\mathcal N=4}^{(SU(2))}}  \frac{1}{2}\oint \frac{db}{2\pi i b}\, Z_{\text{1-loop}}^{\text{VM}}(\mathfrak b, \tau)\,Z^{\text{HM}}(\mathfrak a,\mathfrak b,\tau)\, \big\langle  \tr\left(-\frac{1}{2\ell} Q^i D_z^{\mathfrak A} Q^j \epsilon_{ij} - \tilde\lambda_z \lambda_z \right) \big\rangle_\text{GT}\;.
\end{equation}
The Gaussian expectation value $\langle \ldots \rangle_{\text{GT}}$ can be easily computed using the propagators \eqref{fermionicprop} and \eqref{bosonicprop}. More in detail, as in subsection \ref{HMoneloopcomputation}, we first covariantly point-split the composite operators and then send the separation distance back to zero after having removed the singular terms. The result is
\begin{equation}
\langle T_\text{2d} \rangle_\text{GT} = \frac{1}{8 \pi^2 \ell^2} \bigg[\frac{2 \theta_1''(2\mathfrak b|\tau)}{\theta_1(2\mathfrak b|\tau)} + \frac{\theta_1'''(0|\tau)}{\theta'_1(0|\tau)}  - \sum_{n=-1}^{+1}\frac{\theta_4''(2n\mathfrak b + \mathfrak a|\tau)}{\theta_4(2n\mathfrak b + \mathfrak a|\tau)} \bigg] \; ,
\end{equation}
where $\pm$ means summation over the two signs. Substituting back in \eqref{T2d1pt}, we can perform the contour integrals explicitly, most easily in a series expansion in $q$. Interpreted as the vacuum character of the chiral algebra, it is clear that the $\tau$-derivative of $I_{\mathcal N=4}^{(SU(2))}(q;a)$ is related to the integrated one-point function of the stress tensor. In other words, we expect (the prefactor of the right-hand side of the first equality is obtained similarly to \eqref{comparebefore})
\begin{equation}
\frac{\partial}{\partial \tau} \ln I_{\mathcal N=4}^{(SU(2))}(q;a) =  \frac{\vol(T^2_{\tau})}{\pi(\tau - \bar \tau)}  \langle T_\text{2d} \rangle \qquad \Longleftrightarrow \qquad q \frac{\partial}{\partial q} \ln I_{\mathcal N=4}^{(SU(2))}(q;a)  + \ell^2 \langle T_\text{2d} \rangle =0\;.
\end{equation}
This expectation can be verified analytically by swapping the order of the derivation and integration in $\partial_\tau \ln I_{\mathcal N=4}^{(SU(2))}(q;a) =I_{\mathcal N=4}^{(SU(2))}(q;a)^{-1}\partial_\tau I_{\mathcal N=4}^{(SU(2))}(q;a)$, and is ultimately a consequence of the behavior of the partition function of the symplectic bosons and the $(\lambda_z,\tilde\lambda_z)$-system under $\tau$-derivation, as studied above.

Our next example is the vacuum expectation value of the flavor current $J^{(ij)}$. Obviously only the Cartan component of the current has a nontrivial one-point function. We find
\begin{equation}
\langle J^{12}\rangle =  \frac{1}{I_{\mathcal N=4}^{(SU(2))}} \frac{1}{2}\oint \frac{db}{2\pi i b}\, Z_{\text{1-loop}}^{\text{VM}}(\mathfrak b, \tau)\,Z^{\text{HM}}(\mathfrak a,\mathfrak b,\tau)\, \frac{1}{4\pi \ell}\bigg[\sum_{n=-1}^{+1}\frac{\theta'_4(2n\mathfrak b + \mathfrak a)}{\theta_4( 2n\mathfrak b+\mathfrak a)}  \bigg].
\end{equation}
This result can be related to the $\mathfrak a$-derivative of $I_{\mathcal N=4}^{(SU(2))}(q;a)$. Indeed, we find
\begin{equation}
\frac{\partial}{\partial \mathfrak a} \ln I_{\mathcal N=4}^{(SU(2))}(q;a) =  -4\pi \ell \langle J^{12}\rangle\;.
\end{equation}
We have verified this equality in a series expansion in $q$.

Let us now consider the one-point function of the Sugawara stress tensor $T_{\text{Sug.}}$. It is computed by the contour integral
\begin{align} \nn
&\langle J^{i}{_j}J_i{^j} \rangle \\ \nn
& = \frac{1}{I_{\mathcal N=4}^{(SU(2))}} \frac{1}{2} \oint \frac{db}{2\pi i b}\, Z_{\text{1-loop}}^{\text{VM}}(\mathfrak b, \tau)\,Z^{\text{HM}}(\mathfrak a,\mathfrak b,\tau)\,  \frac{1}{8\pi^2 \ell^2}\bigg[
  \frac{3 \theta'''_1(0)}{\theta'_1(0)}
  + \frac{2\theta'_4(-2\mathfrak b+\mathfrak a)^2}{\theta_4(-2\mathfrak b+\mathfrak a)^2}
  + \frac{2\theta_4'(2\mathfrak b+\mathfrak a)^2}{\theta_4(2\mathfrak b+\mathfrak a)^2}\\
  & + \frac{2 \theta_4'(-2\mathfrak b + \mathfrak a)\theta_4'(\mathfrak a)}{\theta_4(-2\mathfrak b + \mathfrak a)\theta_4(\mathfrak a)} 
   - \frac{2\theta_4'(-2\mathfrak b+\mathfrak a)\theta_4'(2\mathfrak b+\mathfrak a)}{\theta_4(-2\mathfrak b+\mathfrak a)\theta_4(2\mathfrak b+\mathfrak a)}
  + \frac{2\theta_4'(\mathfrak a) \theta_4'(2\mathfrak b+\mathfrak a)}{\theta_4(\mathfrak a) \theta_4(2\mathfrak b+\mathfrak a)}  
  - \sum_{n = -1}^{+1}\frac{3\theta_4''(2n\mathfrak a + \mathfrak b)}{\theta_4(2n\mathfrak a+\mathfrak b)}
  \bigg] \; .
\end{align}
Given that the Sugawara stress tensor is identified with the stress tensor $T_{\text{2d}}$, we should in particular be able to verify that
\begin{align}
\langle T_\text{2d} \rangle = \langle J^{i}{_j}J_i{^j} \rangle \;.
\end{align}
In a series expansion in $q$, it is indeed easy to confirm the equality of the left-hand and the right-hand side.


\section{Discussion and future directions}\label{section:discussion}
In this paper, we have performed a novel localization computation of the Schur limit of the superconformal index of Lagrangian four-dimensional $\mathcal N=2$ SCFTs with the aim of clarifying and elucidating the equality of this particular limit of the index to the (graded) torus partition function of the chiral algebra associated to the four-dimensional theory \`{a} la \cite{Beem:2013sza}. The novelty resides in the choice of localizing supercharge $\mathcal Q$. We found that the hypermultiplets localize to gauged symplectic boson systems living on a torus $T^2$ inside $S^3\times S^1$, while the vector multiplet simply localizes to flat connections. 

The supertrace over (the vacuum module of) the chiral algebra can naturally be further enriched with the insertions of chiral algebra fields. Similarly, we showed in detail that the localization computation of the four-dimensional theory on $S^3\times S^1$ can be applied to compute the corresponding correlators. To do so, we extended the standard lore of what type of insertions can be localized in two directions. First, we showed that the insertion does not necessarily have to be (off-shell) $\mathcal Q$-closed to be localizable. Second, we showed that there is no obstruction to localizing fermionic observables (as long as they are localizable). The result takes the form of a matrix integral coupled to a Gaussian theory, which defines the dynamics of the letters that can be used to string localizable words. The Gaussian theory describing the relevant letters of the hypermultiplets is of course the above-mentioned symplectic boson theory, and the propagator of the relevant letters of the vector multiplet -- certain linear combinations of the gaugini -- implicitly defines the small $(b,c)$ ghost system.

The computation in this paper open various future directions. Let us mention a few particularly interesting ones:
\begin{itemize}
\item First of all, it would be very interesting to analyze existing localization computations to verify if they also admit the insertion of a larger collection of (fermionic) local operators. Of course, we expect the four-sphere analog of the computation in this paper (see \cite{Pan:2017zie}) to similarly admit many (fermionic) insertions, but other computations may also allow for an overlooked, richer set of insertions.
\item In this paper we have only considered the insertion of local operators. The chiral algebra construction in flat space can be generalized by inserting surface operators. To preserve the supercharges $\qq_i$, the defects should be chosen perpendicular to the plane in which the chiral algebra is defined. Their presence gives rise to a module of the vertex operator algebra \cite{Beem:unpubl,Cordova:2016uwk,Cordova:2017mhb,Neitzke:2017cxz,Pan:2017zie,Nishinaka:2018zwq}. Similarly, we can dress our localization computation with defects. In the absence of local operators, this setup computes the (super)character of the associated module. Additional local operator insertions correspond to the obvious insertions in the supertrace over the module.
\item The (unflavored) vacuum character of the vertex operator algebra associated to four-dimensional $\mathcal N=2$ SCFTs satisfies a linear modular differential equation, see \cite{Arakawa:2016hkg,Beem:2017ooy}.\footnote{These differential equations can also be flavored, see \cite{Beem:unpubl2}.} The solutions to such modular differential equations are organized in a vector-valued modular form. In other words, the modular transformation of the vacuum character is a linear combination of the solutions to the equation. The non-vacuum solutions are characters of modules that (conjecturally) correspond to a collection of ``nice'' defects. It is an outstanding problem to analyze the modular behavior of the $S^3\times S^1$ partition function directly and to find out how the ``nice'' defects emerge.
\item As mentioned in footnote \ref{othercorrespondences}, three-dimensional $\mathcal N=4$ SCFTs participate in a correspondence with deformation quantizations \cite{Chester:2014mea,Beem:2016cbd}. If the three-dimensional theory is obtained as a dimensional reduction of a four-dimensional theory, it seems natural to expect that the chiral algebra may be related to said deformation quantization. Our localization setup on $S^3\times S^1$ seems to provide the perfect setup to analyze such relation, as it is well-suited to perform a circle reduction and to make contact with the work in \cite{Dedushenko:2016jxl,Dedushenko:2017avn,Dedushenko:2018icp}, that computes the deformation quantization from results on the three-sphere.
\end{itemize}
We plan to report on some of these directions in the future.

\section*{Acknowledgments}
The authors would like to thank Chris Beem, Bruno Le Floch, Madalena Lemos and Carlo Meneghelli for useful discussions and/or helpful suggestions. Y.P. is supported in part by the 100 Talents Program of Sun Yat-sen University under Grant No.74130-18831116. The work of W.P. is partially supported by grant \#{}494786 from the Simons Foundation.

\appendix

\section{Special Functions}\label{specialfuctions}
In this appendix we collect definitions and a few useful facts regarding various special functions, see for example \cite{bruinier20081}. We define the eta function $\eta(\tau)$ in the standard manner as
\begin{equation}
\eta(\tau) \colonequals q^{1/24} \prod_{n=1}^{\infty}(1-q^n)\;,
\end{equation}
with $\tau\in\mathbb H^+$, \ie{}, $\tau$ takes value in the upper half-plane, and
\begin{equation}
q\colonequals e^{2\pi i \tau}\;.
\end{equation} 
We define the theta function with characteristics as
\begin{equation}
\theta_{\alpha\beta}(z|\tau)\colonequals \sum_{k\in\mathbb Z} \exp\bigg[\pi i \big(k+\frac{\alpha}{2}\big)^2\tau \,+\, 2\pi i \big(k+\frac{\alpha}{2}\big)\big(z+\frac{\beta}{2}\big)\bigg]\;,
\end{equation}
where $z$ is a complex variable. For the purposes of this paper, it is often sufficient to restrict attention to integer-valued characteristics $\alpha, \beta$. Note that real shifts of the argument $z$ can be compensated for by a non-integer characteristic $\beta$. In particular, we have
\begin{align}
-\theta_{11}(z|\tau)&\equiv \theta_1(z|\tau) = -i \sum_{r\in \mathbb Z+\frac{1}{2}} (-1)^{r-\frac{1}{2}}\zeta^r q^{\frac{r^2}{2}} = -i \zeta^{\frac{1}{2}} q^{\frac{1}{8}} \prod_{n=0}^\infty (1-q^{n+1})(1-\zeta q^{n+1})(1-\zeta^{-1}q^n)\nonumber\\
\theta_{10}(z|\tau)&\equiv\theta_2(z|\tau) = \sum_{r\in \mathbb Z+\frac{1}{2}} \zeta^r q^{\frac{r^2}{2}} = \zeta^{\frac{1}{2}} q^{\frac{1}{8}} \prod_{n=0}^\infty (1-q^{n+1})(1+\zeta q^{n+1})(1+\zeta^{-1}q^n)\nonumber\\
\theta_{00}(z|\tau)&\equiv\theta_3(z|\tau) = \sum_{n\in \mathbb Z} \zeta^n q^{\frac{n^2}{2}}=\prod_{n=0}^\infty (1-q^{n+1})(1+\zeta q^{n+\frac{1}{2}})(1+\zeta^{-1}q^{n+\frac{1}{2}})\\
\theta_{01}(z|\tau)&\equiv\theta_4(z|\tau) = \sum_{n\in \mathbb Z} (-1)^n \zeta^n q^{\frac{n^2}{2}}=\prod_{n=0}^\infty (1-q^{n+1})(1-\zeta q^{n+\frac{1}{2}})(1-\zeta^{-1}q^{n+\frac{1}{2}})\;,\nonumber\
\end{align}
where $\zeta\colonequals e^{2\pi i z}$. The expressions in terms of products can be derived using Jacobi's triple product. Note that the fact that
\begin{equation}
\theta_{\alpha+2m,\beta+2n} = (-1)^{\alpha n} \theta_{\alpha\beta}\;,\qquad \text{for }m,n\in\mathbb Z\;,
\end{equation}
implies that the characteristics $(00),(01),(01)$ and $(11)$ are the only four independent (integer) choices. 

The theta functions satisfy the periodicity conditions
\begin{equation}
\theta_{\alpha\beta}(z+n+m\tau|\tau) = (-1)^{n\alpha-m\beta} e^{-\pi i m(2z+m\tau)} \theta_{\alpha\beta}(z|\tau)\;,
\end{equation}
and obey the differential equation
\begin{equation}
4\pi i \frac{\partial \theta_{\alpha\beta}}{\partial\tau} = \frac{\partial^2 \theta_{\alpha\beta}}{\partial z^2}\;.
\end{equation}

Let us also define the twisted Weierstrass functions $P_{k \ge 1}\left[\begin{smallmatrix} \psi \\ \phi \end{smallmatrix}\right](z|\tau)$ as the infinite sums \cite{tuite2007torus}
\begin{equation}
P_k \left[\begin{smallmatrix} \psi \\ \phi \end{smallmatrix}\right](z|\tau) \colonequals \sum_{m, n = - \infty}^{+\infty} \frac{\phi^{-m} \psi^{-n}}{(z + 2\pi (m + n \tau))^k} \; ,
\end{equation}
It was shown in \cite{tuite2007torus} that they satisfy the recursion relation
\begin{equation}
P_k\left[\begin{smallmatrix} \psi \\ \phi \end{smallmatrix}\right](z|\tau) = \frac{(-1)^{k}i}{(k-1)!} \partial_z^{k-1}P_1\left[\begin{smallmatrix} \psi \\ \phi \end{smallmatrix}\right](z|\tau) \; .
\end{equation}
The seed of this recursion relation can be written in terms of Jacobi theta functions as \cite{tuite2007torus}
\begin{equation}
P_1\left[\begin{smallmatrix} \psi \\ \phi \end{smallmatrix}\right](z|\tau)
  = -\eta(\tau)^3 \frac{ \theta_{2\lambda+1,2\mu+1}(\frac{z}{2\pi}|\tau) }{ \theta_{2\lambda+1,2\mu+1}(0|\tau) \ \theta_{11}(\frac{z}{2\pi}|\tau) } \ ,
\end{equation}
for $\phi = e^{2\pi i \lambda}$, $\psi = e^{- 2\pi i \mu}$.


\section{The Lagrangian supermultiplet and \texorpdfstring{$\mathfrak{Q}$}{Q}-exactness \label{app:Q-exactness}}
In this appendix we collect some details on the $\mathcal N=2$ (anti)chiral multiplet whose top component is the (anti)holomorphic part of the Yang-Mills Lagrangian. While it can be constructed as an abstract multiplet, it will be necessary for us to have an off-shell realization in terms of the elementary fields of the vector multiplet, \ie{}, $A_\mu, \phi, \tilde \phi, \lambda_I, \tilde \lambda_I$, and $D_{IJ}$. From these (adjoint-valued) fields we construct the following composite objects,
\begin{align}
&\Phi \colonequals \frac{1}{2}\tr\phi^2\;, \qquad B_{IJ} \colonequals \tr \left(\phi D_{IJ} + \frac{i}{2} (\lambda_I \lambda_J)\right)\;, \qquad \; \notag G_{\mu \nu} \colonequals \tr\left( \phi F_{\mu \nu} + \frac{i}{8}(\lambda^I \sigma_{\mu \nu } \lambda_I)\right)\\
& \Psi_I \colonequals -i \tr \phi \lambda_I\;, \qquad \Lambda_I \colonequals \tr \left(- i F_{\mu \nu}\sigma^{\mu \nu} \lambda_I + 4i  \phi \sigma^\mu D_\mu \tilde\lambda_I - 2i D_I{^J} \lambda_J +4 [\phi,\tilde\phi]\lambda_I\right)\;,
\end{align}
and a similar set of fields $(\tilde \Phi, \tilde \Psi, \tilde \Lambda_I, \tilde B_{IJ}, \tilde G_{\mu\nu})$ with $\phi \to \tilde \phi$, $\lambda \to \tilde \lambda_I$, along with suitable $\sigma$-matrices replacement. We also consider the holomorphic and anti-holomorphic part of the full Yang-Mills Lagrangian
\begin{align}
& \mathcal{L}^-_\text{YM} \colonequals \mathcal{L}_\text{YM} + 4 \tr D_\mu \big[  \phi D^\mu \tilde \phi  \big] - \frac{1}{4}\epsilon^{\mu \nu \lambda \rho} \tr F_{\mu \nu} F_{\lambda \rho} \\
& \mathcal{L}^+_\text{YM} \colonequals \mathcal{L}_\text{YM} + 4 \tr D_\mu \big[  \tilde \phi D^\mu \phi  \big] + \frac{1}{4}\epsilon^{\mu \nu \lambda \rho} \tr F_{\mu \nu} F_{\lambda \rho}\ ,
\end{align}
where $\mathcal{L}_\text{YM}$ was given in \eqref{VM-Lagr}. The full supersymmetric Yang-Mills action with $\theta$-term on a manifold $M$ is then given as
\begin{equation}
S_{\text{YM}} = \frac{1}{g_{\text{YM}}^2}\int_M d^4x \sqrt{g}\ \mathcal{L}_{\text{YM}} + \frac{i\theta}{8\pi^2}\int_M \tr{} F\wedge F \nn = \frac{1}{8 \pi i} \int_M d^4x \sqrt{g}\ \left(\tau_{\text{YM}} \mathcal{L}^-_\text{YM} - \bar\tau_{\text{YM}} \mathcal{L}^+_\text{YM}\right)\;,
\end{equation}
where $\tau_{\text{YM}} = \frac{\theta}{2\pi} + \frac{4\pi i}{g_{\text{YM}}^2}$ is the exactly marginal coupling.

From \eqref{VM-SUSY}, it is straightforward to organize $( \Phi,  \Psi,  \Lambda_I,  B_{IJ},  G_{\mu\nu}, \mathcal{L}_\text{YM}^-)$ into a chiral supermultiplet:
\begin{align*}
  \delta \Phi = &\ (\xi^I \Psi_I) \\
  \delta \Psi_I = &\  - \frac{i}{2} G_{\mu \nu} \sigma^{\mu \nu} \xi_I - i B_{IJ} \xi^J - 2i \sigma^\mu D_\mu (\Phi \tilde \xi_I) \\
  \delta B_{IJ} = &\ \Big[(\tilde \xi_I \tilde \sigma D_\mu \Psi_J) + 2i (\xi'_I \Psi_J) + \frac{1}{4}(\xi_I \Lambda_J) \Big] + (I \leftrightarrow J)\\
  \delta G_{\mu \nu} = & \ -\frac{1}{8} (\xi^I \sigma_{\mu \nu} \Lambda_I) + \frac{1}{2} D_\lambda(\tilde \xi^I \sigma^\lambda \sigma_{\mu \nu} \Psi_I) + i (\xi'^I \sigma_{\mu\nu} \Psi_I) \\
  \delta \Lambda_I = & \ i G_{\mu \nu} \sigma^{\mu \nu} \sigma^\lambda D_\lambda \tilde \xi_I - 2i D_\rho \big[G_{\mu \nu} \sigma^{\mu \nu} \sigma^\rho \tilde \xi_I \big] + 2i \mathcal{L^-_\text{YM}} \xi_I  \\
   & \ + 4i (D_\mu B_{IJ}) \sigma^\mu \tilde \xi^J + 6i B_{IJ} \sigma^\mu D_\mu \tilde \xi^J \;. \\
   \delta \mathcal{L^-_\text{YM}} = & D_\mu (\tilde\xi^I \tilde \sigma^\mu \Lambda_I) \ .
\end{align*}
Here the spinors $\xi$ and $\tilde \xi$ satisfy the (generalized) Killing spinor equations (\ref{KSE}) and (\ref{auxKSE}). Recall that the most general solution to these equations on our supergravity background was presented in \eqref{genericQ}. Similarly, one can construct the antichiral multiplet $(\tilde \Phi, \tilde \Psi, \tilde \Lambda_I, \tilde B_{IJ}, \tilde G_{\mu\nu}, \mathcal L_{\text{SYM}}^+)$. 

To show that the Yang-Mills action is supersymmetrically exact on our supergravity background, we define
\begin{align}
  \Xi^- \colonequals - \frac{\ell}{c_1 c_2} (\xi'^I \Lambda_I) \ ,
  \qquad
  \Xi^+ \colonequals - \frac{\ell}{c_1 c_2} (\tilde\xi'^I \tilde\Lambda_I) \ ,
\end{align}
It is then straightforward, though somewhat tedious, to show that
\begin{align}\label{Q-exactness-of-YM}
  \int_{S^3\times S^1} \sqrt{g} \mathcal{L}_\text{YM}^\pm = \delta_{\xi, \tilde \xi} \int_{S^3\times S^1} \sqrt{g} d^4 x \tr \Xi^\pm \ .
\end{align}
In particular,
\begin{align}
  S_\text{YM} = \frac{1}{8 \pi i} \delta_{\xi, \tilde \xi}\bigg[\int \sqrt{g}d^4 x \left(\tau_\text{YM} \Xi^- - \bar \tau_\text{YM} \Xi^+ \right)\bigg] \ .
\end{align}

It will be useful to slightly rewrite the terms containing auxiliary fields in the composite $\Xi^{\pm}$, \ie{}, the terms to $ - \frac{2i \ell}{c_1 c_2} D_{IJ} (\xi'^I\lambda^J)$ and $- \frac{2i \ell}{c_1 c_2}  D_{IJ} (\tilde\xi'^I\tilde\lambda^J)$. A useful identity for this purpose is, for two generic symmetric tensors $X_{IJ}$ and $Y_{IJ}$,
\begin{align}\label{fierz}
  \frac{1}{2}X^{IJ}Y_{IJ} = & \ + \frac{1}{2}\Big[X_{IJ}(R^{IJ}_4 + i R^{IJ}_1) \Big] \Big[ \frac{1}{s \tilde s} Y_{KL}(R^{KL}_4 - i R^{KL}_1) \Big] \\
    & \ + \frac{1}{2}\Big[\frac{1}{s \tilde s}X_{IJ}(R^{IJ}_4 - i R^{IJ}_1) \Big] \Big[  Y_{KL}(R^{KL}_4 + i R^{KL}_1) \Big] + \Big[ \frac{1}{\sqrt{s \tilde s}} X_{IJ}R^{IJ}_3  \Big]\Big[ \frac{1}{\sqrt{s \tilde s}} Y_{IJ}R^{IJ}_3  \Big] \nonumber \ .
\end{align}
Among the resulting terms when applied to $ - \frac{2i \ell}{c_1 c_2} D_{IJ} (\xi'^I\lambda^J)$, there is a particular interesting one of the form
\begin{equation}\label{Xiz}
\Big[\frac{2i}{\ell s\tilde s}D^{IJ}(R_{IJ}^4 - i R_{IJ}^1)\Big] \Big[(2i)\frac{\ell}{2i}\Big(  - \frac{\ell}{c_1 c_2}  \Big)(\xi'_I \lambda_J)  (R^{IJ}_4 + i R^{IJ}_1) \Big] = - \cos^2\frac{\vartheta}{2}\mathfrak{D}_{\bar z}\lambda_z\;,
\end{equation}
and its $\tilde \lambda$ cousin. Here $\lambda_z \colonequals \frac{\ell}{2i} (\tilde \xi^I (\tilde \sigma_4 + i \tilde \sigma_1)\lambda_I)$ at a generic point $x$, and similarly for $\tilde \lambda_z$.

\clearpage

{
\bibliographystyle{utphys}
\bibliography{ref}
}

\end{document}